\documentclass[]{elsarticle} 
\usepackage[hyphens]{url}

  \journal{Journal of Artificial Societies and Social Simulation} 

\usepackage{lineno} 
\providecommand{\tightlist}{%
  \setlength{\itemsep}{0pt}\setlength{\parskip}{0pt}}

\usepackage{graphicx}
\usepackage{booktabs} 

\usepackage[T1]{fontenc}
\usepackage{lmodern}
\usepackage{amssymb,amsmath}
\usepackage{ifxetex,ifluatex}
\usepackage{fixltx2e} 
\IfFileExists{upquote.sty}{\usepackage{upquote}}{}
\ifnum 0\ifxetex 1\fi\ifluatex 1\fi=0 
  \usepackage[utf8]{inputenc}
\else 
  \usepackage{fontspec}
  \ifxetex
    \usepackage{xltxtra,xunicode}
  \fi
  \defaultfontfeatures{Mapping=tex-text,Scale=MatchLowercase}
  
\fi
\IfFileExists{microtype.sty}{\usepackage{microtype}}{}
\usepackage{longtable}
\usepackage{graphicx}
\makeatletter
\def\maxwidth{\ifdim\Gin@nat@width>\linewidth\linewidth
\else\Gin@nat@width\fi}
\makeatother
\let\Oldincludegraphics\includegraphics
\renewcommand{\includegraphics}[1]{\Oldincludegraphics[width=\maxwidth]{#1}}
\ifxetex
  \usepackage[setpagesize=false, 
              unicode=false, 
              xetex]{hyperref}
\else
  \usepackage[unicode=true]{hyperref}
\fi
\hypersetup{breaklinks=true,
            bookmarks=true,
            pdfauthor={},
            pdftitle={A Demographic Microsimulation Model with a household alignment method.},
            colorlinks=false,
            urlcolor=blue,
            linkcolor=magenta,
            pdfborder={0 0 0}}
\newlength{\cslhangindent}
\newenvironment{cslreferences}%
  {\setlength{\parindent}{0pt}%
  \everypar{\setlength{\hangindent}{\cslhangindent}}\ignorespaces}%
  {\par}

\usepackage{pdflscape}
\newcommand{\blandscape}{\begin{landscape}}
\newcommand{\elandscape}{\end{landscape}}
\usepackage[ruled,vlined,linesnumbered,boxed]{algorithm2e}
\usepackage[section]{placeins}
\usepackage{float}
\usepackage{algpseudocode}
\usepackage{mathptmx,amsmath}
\usepackage{booktabs}
\usepackage{longtable}
\usepackage{array}
\usepackage{multirow}
\usepackage{wrapfig}
\usepackage{float}
\usepackage{colortbl}
\usepackage{pdflscape}
\usepackage{tabu}
\usepackage{threeparttable}
\usepackage{threeparttablex}
\usepackage[normalem]{ulem}
\usepackage{makecell}
\usepackage{xcolor}

\begin{document}
\begin{frontmatter}

  \title{A demographic microsimulation model with an integrated household alignment method.}
    \author[UNSW Sydney]{Amarin Siripanich}
   \ead{a.siripanich@unsw.edu.au} 
    \author[UNSW Sydney]{Taha Rashidi\corref{1}}
   \ead{rashidi@unsw.edu.au} 
      \address[UNSW Sydney]{UNSW Sydney, NSW, 2052, Australia}
      \cortext[1]{Corresponding Author}
  
  \begin{abstract}
  Many dynamic microsimulation models have shown their ability to reasonably project detailed population and households using non-data based household formation and dissolution rules. Although, those rules allow modellers to simplify changes in the household construction, they typically fall short in replicating household projections or if applied retrospectively the observed household numbers. Consequently, such models with biased estimation for household size and other household related attributes lose their usefulness in applications that are sensitive to household size, such as in travel demand and housing demand modelling. Nonetheless, these demographic microsimulation models with their associated shortcomings have been commonly used to assess various planning policies which can result in misleading judgements. In this paper, we contribute to the literature of population microsimulation by introducing a fully integrated system of models for different life event where a household alignment method adjusts household size distribution to closely align with any given target distribution. Furthermore, some demographic events that are generally difficult to model, such as incorporating immigrant families into a population, can be included. We illustrated an example of the household alignment method and put it to test in a dynamic microsimulation model that we developed using dymiumCore, a general-purpose microsimulation toolkit in R, to show potential improvements and weaknesses of the method. The implementation of this model has been made publicly available on GitHub, a code sharing platform.
  \end{abstract}
 \end{frontmatter}

\hypertarget{introduction}{%
\section{Introduction}\label{introduction}}

Models that can project human populations at the levels of individual and household are a starting point modelling complex systems. Their explore people's behaviour and interactions, as they advance through different stages of their lives and as their environment -- such as policies, social trends and the economy -- changes. This modelling method is known as dynamic microsimulation, which is a subset of microsimulation models which changes over time. It has been widely adopted by many studies where agents' behaviour and their states are dynamic with their life courses, such as economic decisions (Fatmi and Habib, 2018), disability status (van Sonsbeek and Alblas, 2012), and health outcomes (Rutter et al., 2011). It has also gained recognition as an attractive tool for policy analysis (Figari et al., 2015). The amount of insights that can be generated by a dynamic microsimulation model enables impact assessment under different policy scenarios across multiple interesting dimensions, such as, in different groups of population and also spatio-temporal scales (Zaidi and Rake, 2001).

Several demographic microsimulation methods have shown their ability to produce simulated populations matching some characteristics of the real populations, by using a set of simple household formation rules and demographic transition models. Although, those rules are often based on no empirical data, such as assuming that adult leavers will create one-person households (Galler, 1988), they can still produce an acceptable result for short-period validation, as changes in the demographic structure of most populations are observed in a long run. Therefore, the effectiveness of these rules might be questionable as their performance has not been examined versus not having any rules in long-run scenarios. One very important characteristic of future populations is the distribution of their household sizes. It is a significant factor in forecasting travel demand, housing demand, and household expenditures. An obvious resolution would be to develop behavioural models to simulate life events. However, this may not be possible for many studies where appropriate household data is missing over time to observe the dynamics of household evolution. Another alternative would be to align the micro-level outcomes to their the target projections which can be transferred to future with the aim of maintaining the overall structure of households on the population.

In this study, we present a dynamic microsimulation model for projecting future structure of population and households with a novel alignment method to allocate newly formed and newly immigrated households to the population such that it closely matches pre-defined household size distributions that can vary over time.  We developed `dymiumCore'\footnote{www.github.com/dymium-org/dymiumCore}, a general-purpose microsimulation toolkit in R which incorporate several individual and household level decisions. The modular structure of our toolkit facilities its maintains as more behavioural models are introduced to the toolkit to improve the accuracy of the whole microsimulation platform. The platform is unique of its kind in terms of its holistic structure of lifestyle events as well as the efficiency of the algorithms developed that can efficiently update lifestyle changes for all agents defined in the entire populations. 

The rest of the paper is structured as the following. The next section presents the literature review and background of the study, follows by the discussion of the proposed model, data, assumptions, and sub-models. The model was run multiple times and those results were used for validation. Lastly, we conclude lessons learnt from this study and possible future improvements.

\hypertarget{literature-review-and-background}{%
\section{Literature review and background}\label{literature-review-and-background}}

\hypertarget{microsimulation}{%
\subsection{Microsimulation}\label{microsimulation}}

The term \emph{microsimulation} has been given slightly different meanings depending on which context it appears in. However, the board meaning of the term refers to a computerised micro-analytical approach that generates stochastic micro-level and highly complex outcomes which emerge from interactions between agents (e.g.~people, firms, and vehicles) and their realisation, such as a state transition. Hence this approach requires highly detailed unit record data and parameters that describe behaviour of their agents and environment. In this paper, we refer to microsimulation in the context of econometric and social science, as pioneered by Orcutt (1957) in the late 1950s.

The earlier work of Guy Orcutt provides the foundation for microanalytic simulation approaches. This highly influential work paved a promising avenue for demographers, economists and social scientists to breakaway from traditional aggregate approaches for forecasting population, such as the `headship rate' approach, to account for the effects of individual behaviour (Galler, 1988). Clarke (1986) provides a review of the early development of population and household microsimulation models. Microsimulation can be used not only to model evolving populations and their relationships, but it is also applicable for simulating other decisions and changes in characteristics and behaviour through out their life-courses. This type of microsimulation models are referred to as a dynamic microsimulation model.

Many dynamic microsimulation models have been developed for policy evaluations at national level across the world. Some examples are MOSART for Norway's labour supply and public pension benefits (Fredriksen, 1998), APPSIM for many policy-related questions specific to the future Australian population (Harding, 2007), INAHSIM for Japan's household living arrangement and public assistance related policies (Fukawa, 2011). SESIM for Sweden's ageing population's impact on their pension system and public finances (Flood et al., 2005), POLISIM for USA's social security projection (McKay, 2003), LIAM for evaluating reform scenarios to the Irish pension system (O'Donoghue et al., 2009), MIDAS for analysing the social security and pension systems of Belgium, Germany and Italy (Dekkers et al., 2010).

As a multidisciplinary tool, several use cases exist for microsimulation platforms, such as in health care (Zucchelli et al., 2010), household formation and dissolution (Galler, 1988), transportation and land use (Salvini and Miller, 2005), travel demand (Goulias and Kitamura, 1992), housing choice (Benenson and Torrens, 2004), labour market (Harmon and Miller, 2018), and firm life-cycle (also known as `firmography') (Bodenmann, 2011). These social and economic systems rely on an evolving population to simulate dynamic outcomes, by simulating the progression in one's life trajectory.

\hypertarget{projecting-population-and-households}{%
\subsection{Projecting population and households}\label{projecting-population-and-households}}

Many efforts have been put toward development of demographic components of dynamic microsimulation models, specifically to capture population changes at the individual level (person, family, and household) with more behavioural realism (Li and O'Donoghue, 2013). Different models may require different characteristics of their decision-making units, that usually depends on the intended purposes of each model. Models that are intended for examining the spread of long-term infectious diseases (Geard et al., 2013), residential location choice (Moeckel, 2016), and household expenditure (Lawson, 2014) would require their demographic components to accurately capture the household structure and composition of their simulated population. As an example, in the case of residential mobility, changes in the characteristics of a household such as the household composition and the number of household members can affect housing decisions of the household (Rashidi, 2015; Rashidi et al., 2011). Despite those differences, the core demographic processes that are responsible for population growth and family formation and dissolution, namely, marriage, divorce, birth, death, leaving parental home and migration can be found in most instances (Morand et al., 2010) with limitations on interaction and realisim of these processes. These life events of people are usually simulated using statistical models (such as logistic regression and decision tree models) or simple transition probability tables estimated from population surveys. There could be a series of decisions and predicted quantities required in one event. As an example, Eluru et al. (2008) apply their birth event only to women aged between 10 to 49, most models have the lower bound starting from 15, and for each woman that is giving birth and the number of newborns and their genders will be also determined in the process. The demographic decisions of an individual commonly affect their family or household members. For instance, when a married couple gets divorced, one of the partners, usually the male partner, will leave the household while their children, if any, will remain in the old household with the female partner.Therefore, simplifications in demographic models can propogate error into the overall strcutre of the model. 

Although changes in family structure and population growth can be modelled quite easily using microsimulation approaches, modelling household formation and dissolution can be quite challenging, especially when population surveys are scarce. An ideal example is SERIVGE, a Swedish microsimulation model, where longitudinal socioeconomic information of the entire population in Sweden from 1985 to 1995 are available along with highly detailed geographical identifier of each household (Rephann and Holm, 2004). However, many models do not have the luxury as SERIVGE, so they restrcit their models to simple household formation and dissolution assumptions. A set of common assumptions, a de facto standard per se, for household formation and dissolution can be found for several lifestyle models, as shown in Table \ref{tab:demographic-model-review}. By definition, a household can be made up of related and unrelated individuals, while a family only consists of people that are related by blood or by marriage or cohabitation. Most demographic models do not consider the formation of households with non related individuals and family households with other individuals explicitly, hence, these types of households do not get reflected in their model. Consequently, the household size and composition distributions of those models will most likely be unaligned with their official projection or historical data. A few exceptions exists, (Paul, 2014) developed a roommate model which groups non-related individuals into group households, and (Inagaki, 2018)'s model captures multi-generation households and people returning to their parental home after relationship breakdown. However, even those that consider do suffer from creating appropriate sizes of households due to no housing constraints being imposed.

Household formation and dissolution assumptions directly affect the composition and size of new households that get formed, resulting from demographic outcomes, as well as changes to existing households. Chingcuanco and Miller (2018) apply a probabilistic model for the child custody decision that entails partnership dissolution events, while most models consider a far simpler rule such as leaving the children in the old household with their mother. Such simplification may lead to an overestimation in the number of lone parents in one gender and could affect policy decisions derived from such model. Another simplification that is often found in the literature is modelling of migration as net migration (see MOSART, DESTINE and CORSIM). The justification for this is often a lack of migrant data and that some regions expect a net positive number of migrants, hence emigration was not explicitly considered. While this assumption would not have a significant impact on the population distributions, such as by age and sex, in regions where the flows of migrants are not high. However, major cities such as Melbourne or Sydney of Australia have been attractive destinations for migrants, hence, not considering emigration explicitly can severely impact the correctness of the microsimulation result. Other approaches have been introduced to deal with a lack of data, such as the Pageant algorithm, an alignment method, proposed by (Ch\a'enard, 2000) for correcting the population structure when some information about the characteristics of individual migrants is known. Although, the algorithm is used by a number of models, such as by Dekkers (2015), it doesn't incorporate new migrant families joining into existing households, which could paint a wrong picture in the final analysis, for instance, if housing demand is a quantity that the model is projecting. This is supported by Deloitte (2011), they forecast that 36\% of new migrant families in Australia will initially be dependent on existing households for housing, hence, migrants do not put immediate pressure on the housing market when they first arrive.

\hypertarget{correction-of-errors}{%
\subsection{Correction of errors}\label{correction-of-errors}}

Parameter calibration and alignment are the main approaches for correcting the results of a microsimulation model to match external totals (Baekgaard, 2002). Both approaches have been widely used and discussed in many studies. As said by Miller (2018) ``.. microsimulation is not a model per se, but rather is a computational structure for the implementation of models of system behavior'', hence, parameter calibration needs to be done both jointly and individually to ensure the system behaviour is correctly reflected in every part of the model. Parameter calibration refers to a procedure for modifying coefficients of an estimated model, usually only the intercept terms are changed to scale the average proportions, such that the model's output closely matches an exogenous target. While alignment approaches, mechanisms that involve selecting micro units to undergo an event such that the total number of events occurrences is consistent with an external total, are often used to make sure that a microsimulation model produces results that are indifferent from exogenous expectations of future events (Li and O'Donoghue, 2013). Although, alignment has its downsides that are quite concerning as discussed by Baekgaard (2002) and Li and O'Donoghue (2014), Anderson (1997) noted that it is a common practice and can be found in most existing dynamic microsimulation models used for policy analysis.

\hypertarget{contributions}{%
\subsection{Contributions}\label{contributions}}

In this paper, we propose a household size alignment method, integrated into a joint system of behavioural household and individual decision models,  which allows for an external household size distribution to be matched. This method allows new migrant families, people leaving their parental home and those leaving their household due to relationship breakdown to join existing households and create new households as required to match the pre-defined household size target. The reason that we are called this an `alignment-like' method is because it can only match the target distribution of household sizes if the new households can be assigned to those bins that are lower or higher than its expected count. More details of the alignment method will be discussed in the next section. We also made the code of the proposed dynamic microsimulation model that, written in the R languages (R Core Team, 2019), available on GitHub.\footnote{www.github.com/asiripanich/dymium-melbourne-model}

\newpage
\begin{landscape}

\begin{table}

\caption{\label{tab:demographic-model-review}A comparison of the household formation and dissolution rules in household microsimulation models.}
\centering
\resizebox{\linewidth}{!}{
\begin{tabu} to \linewidth {>{\centering\arraybackslash}p{8em}>{\raggedright\arraybackslash}p{8em}>{\raggedright\arraybackslash}p{13em}>{\raggedright\arraybackslash}p{13em}>{\raggedright\arraybackslash}p{13em}>{\raggedright\arraybackslash}p{13em}>{\raggedright\arraybackslash}p{5em}}
\toprule
\textbf{Reference} & \textbf{Use Case} & \textbf{Partnership Formation} & \textbf{Partnership Dissolution} & \textbf{Leaving Parental Home} & \textbf{Migration} & \textbf{Code Publicly Availiable}\\
\midrule
\em{Common assumptions} & \em{} & \em{Create a new household where both partners and their dependants  (normally only include resident children) move in,  or one partner joins another household along with dependants.} & \em{The male partner leaves its family household and create a new lone household. Their children usually stay in the same household with the female partner.} & \em{Leavers create new one-person households} & \em{Add new migrant families as new households to the population and often only net migration is considered.} & \em{}\\
\cmidrule{1-7}
Galler (1988) & Household dynamics & Randomly select one of the CAs. &  &  & Did not consider explicitly. & No\\
\cmidrule{1-7}
Rephann and Holm (2004) & Policy evaluation & The female partners move into their males' households after getting married. &  &  & Consider immigration and emigration separately. & \\
\cmidrule{1-7}
Eluru et al. (2008) & Activity-based modelling & Randomly select one of the CAs. & Consider child custody between the parents. & Leavers can form non-family households of various sizes and can leave the study area. & Consider immigration and emigration separately. New migrants can join the population as new households or join existing households. & No\\
\cmidrule{1-7}
Fukawa (2011) & Projection of health and long-term care expenditures & Each of the CAs has a different probability value. & Also consider people returning their parental homes after relationship breakdown. & Doesn't consider children leaving home for reasons other than to get married. & Doesn't consider migration at all. & No\\
\cmidrule{1-7}
Wu and Birkin (2012) & Regional planning &  & Not stated. & Not stated. & Not stated. & No\\
\cmidrule{1-7}
Geard et al. (2013) & Explore patterns of infection and immunity & If both of the partners are living with their parents, then create a new household. If either of the individuals have their own household then, the other partner along with their children joins them in this household. &  &  & Clone existing households. & Yes\\
\cmidrule{1-7}
Lawson and Anderson (2014) & Forecast household expenditures &  & Also consider people returning their parental homes after relationship breakdown. &  & N/A & Yes\\
\cmidrule{1-7}
Paul (2014) & Activity-based modelling & The female partners move into their males' households after getting married. & Consider child custody between the parents. The partner that gains custody of their children stay in the current household. The other partner leaves to form a non-family household. & Leavers can form non-family households of various sizes. & Consider immigration and emigration separately. & No\\
\cmidrule{1-7}
Moeckel (2016) & Housing and transport demand & Create a new two person households. If either of the partners have children then their children will join the new household. &  &  & Consider immigration and emigration separately. & Yes\\
\cmidrule{1-7}
Chingcuanco and Miller (2018) & Urban land use & New couples can merge their households. & Consider child custody between the parents. &  & Consider immigration and emigration separately. 75\% chance to emigrate as a household and 25\% chance for only household heads to emigrate. & Yes\\
\bottomrule
\multicolumn{7}{l}{\textsuperscript{*} CA = Common assumption}\\
\end{tabu}}
\end{table}

\end{landscape}
\newpage

\hypertarget{the-microsimulation-model}{%
\section{The microsimulation model}\label{the-microsimulation-model}}

\hypertarget{model-structure}{%
\subsection{Model structure}\label{model-structure}}

The main objective of our model is to simulate individuals and households that are closely aligned with Australia official population projections by simulating all the `usual' demographic events and the household alignment method. There are 10 demographic events in the following order: ageing, birth, death, marriage, divorce, cohabitation, break up, leaving parental home, emigration, and immigration. The migration events were simulated separately for interregional migrants and overseas migrants. All these events happened sequentially and in discrete time where one simulation cycle was equal to one year of change. The order of events is known to have a profound impact on the model's results. Using a simple example, if we were to simulate the death event before the birth event, we can expect to see less number of births each year, this is because less women would be alive by the time the birth event is to be simulated. However, if we switch their order the other way around, we should expect to see more births each year than the former order. Only a few studies have explored this particular issue, see Dumont et al. (2018). Despite of that, some justification can be made without requiring a thorough experiment, such as putting the divorce and break up events before the marriage and cohabitation events, to make sure that remarried/cohabitation can happen within the same year, which is also what APPSIM does (Bacon and Pennec, 2007).

Apart from the demographic events, two socio-economic characteristics of all the individuals, labour force status and education status, were also updated in each simulation cycle. Where applicable, these attributes were also used as covariates in the demographic models. Hence, any changes in those attributes of individuals will change their chance of undergoing a demographic event. Ideally, this is where one can implement a macro-micro linkage so that demographic decisions at the individual level can be behaviorally responsive to macroeconomic factors.

\begin{figure}
\centering
\includegraphics{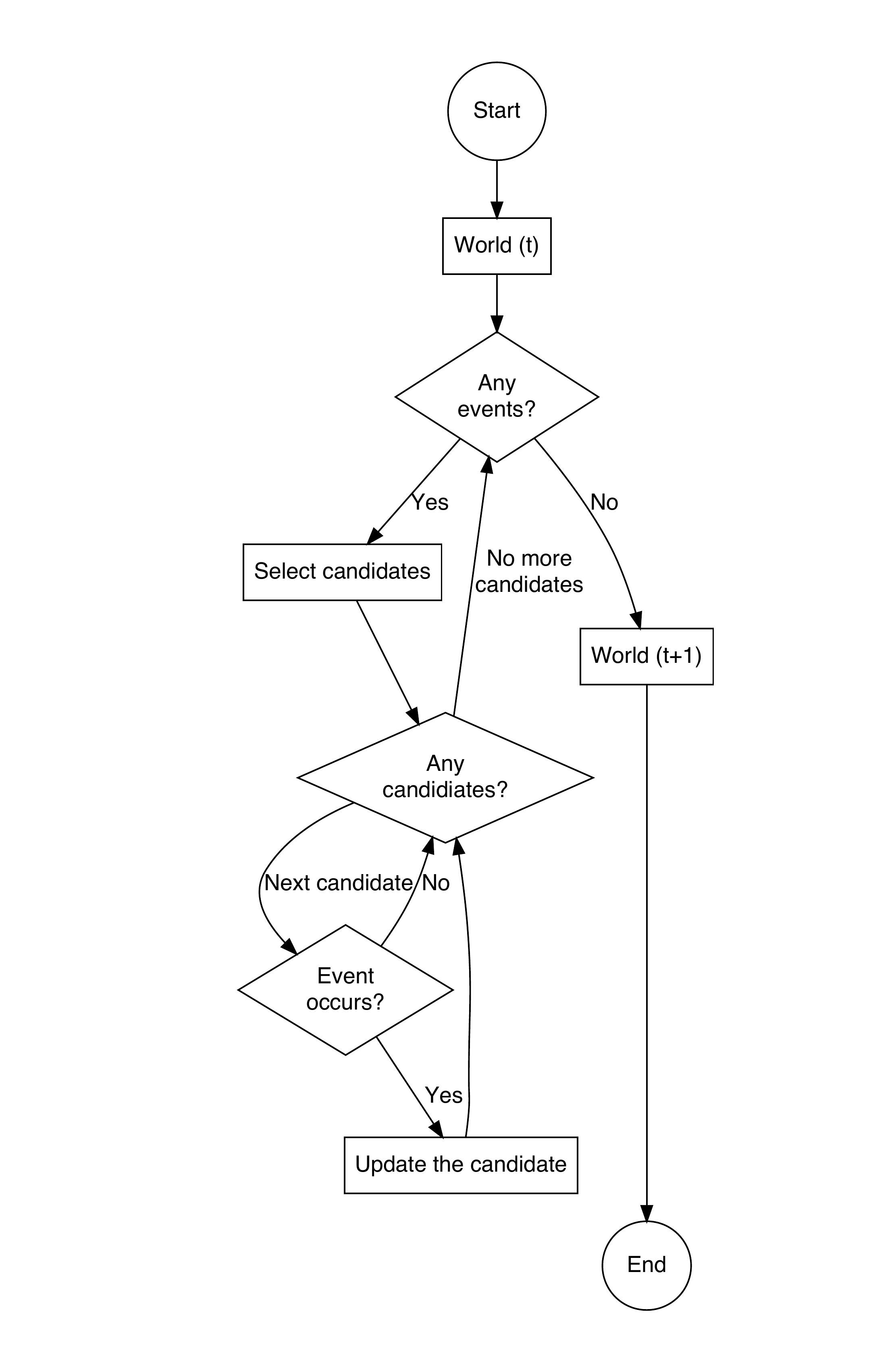}
\caption{\label{fig:microsim-process}A flowchart that shows how a microsimulation model iterates through events in one simulation cycle.}
\end{figure}

Figure \ref{fig:microsim-process} shows a high level picture of how a microsimulation model iterates through events in one simulation cycle. One can think of a microsimulation model as a data pipeline, where a dataset gets passed in and flow through a set of predefined operations, referred to as events, that contains parameters and other settings. However, as this is a stochastic model, there is a chance that not every event will be performed on all of the data points of the input dataset, where each data point represents a unique entity such as a unique person. For an agent to successfully undergo an event (e.g.~to give birth), it depends on the event's candidate selection criteria, deterministic, and the risk probability, stochastic, which usually associated with characteristics of the selected candidates. The stochastic part is simulated using a Monte Carlo simulation. For an agent to undergo an event usually means that some characteristics of that agent will be changed, granted that the event is simulated to occur for that agent. The output of one event will be the input of the next event. This procedure is repeated until all events are done, which marks the end of a simulation cycle.

\hypertarget{initialisation}{%
\subsection{Initialisation}\label{initialisation}}

We used a 1\% basic confidentialised unit record file (also referred to as a microdata set or a reference sample) and tabulations (also known as marginal sums, marginal distributions, and control totals), containing various demographic and socioeconomics characteristics, as shown in Table \ref{tab:synthetic-population}, to synthesise a baseline population for our microsimulation model. The synthetic population file contains records of individuals and their households. All the data used were sourced from the 2011 Australian population and housing census survey. Our baseline population is limited to just the Greater Melbourne region, home to around 4 million residents in 2011, the second largest cities in Australia, and the capital city of the Victoria state.

\begin{table}[!h]

\caption{\label{tab:synthetic-population}Synthetic population and household characteristics}
\centering
\begin{tabu} to \linewidth {>{\bfseries}l>{\raggedright\arraybackslash}p{8em}>{\raggedright\arraybackslash}p{20em}}
\toprule
 & Characteristics & Levels\\
\midrule
\addlinespace[0.3em]
\multicolumn{3}{l}{\textbf{Individual}}\\
\hspace{1em} & Age & 18 categories: 0 - 4, ..., 85+ and over\\
\cmidrule{2-3}
\hspace{1em} & Sex & 2 categories: Male and Female\\
\cmidrule{2-3}
\hspace{1em} & Marital status & 6 categories: Not applicable, Never married, Married, Separated, Divorced, Widowed\\
\cmidrule{2-3}
\hspace{1em} & Employment status & 4 categories: Not applicable, Employed, Unemployed, Not in labour force\\
\cmidrule{2-3}
\hspace{1em} & Student status & 3 categories: Not applicable, Part-time, Full-time\\
\cmidrule{2-3}
\hspace{1em} & Father id & Numeric\\
\cmidrule{2-3}
\hspace{1em} & Mother id & Numeric\\
\cmidrule{2-3}
\hspace{1em} & Partner id & Numeric\\
\cmidrule{1-3}
\addlinespace[0.3em]
\multicolumn{3}{l}{\textbf{Household}}\\
\hspace{1em} & Household size & 6 categories: 1, 2, ..., 6+\\
\cmidrule{2-3}
\hspace{1em} & Place of residence & 1 category: Greater Melbourne\\
\bottomrule
\end{tabu}
\end{table}

To generate a baseline synthetic population, we followed a standard synthetic
reconstruction procedure which involves two stages: fitting and generation
(Müller, 2017).

For the fitting stage, the iterative proportional updating method, also known as IPU, was used. It is a heuristic reweighing approach proposed by Ye et al. (2009). This method is known to be highly efficient and easy to understand, as it is an Iterative Proportional Fitting procedure. The highlight of this method is its ability to calibrate the case weights of a population sample to match their individual-level and household-level marginal sums. Hence, the calibrated weight of each record in the reference sample reflects the record's contribution at both of the levels. There are many software packages that provide implementations of various fitting methods. For this study we used the simPop package's implementation of the IPU method (Templ et al., 2017).

Different demographic models may have different requirements. This decision largely depends on what they are developed for. The specification of synthetic population is one of the many requirements to be discussed prior to the model development phrase. For this study, in addition to demonstrating the effective of the overall microsimulation framework, one of our main goals is to show that the proposed allocation algorithm works as intended. Hence we kept the specification of our synthetic population as simple as possible. The following individual-level cross-tabulations were used as control tables:

\begin{itemize}
\tightlist
\item
  \emph{age} x \emph{sex} x \emph{marital status},
\item
  \emph{age} x \emph{sex} x \emph{employment status}, and
\item
  \emph{age} x \emph{sex} x \emph{student status}.
\end{itemize}

Only household size was used as the household-level constraint.

Once the calibrated weights of the reference sample were obtained, they had to be integerised to determine how many copies of each unique record should be generated to mirror the real population. There are many ways to perform integerisation. We picked the TRS approach as it strikes the balance between the ease of implementation and has a superior accuracy than other approaches as shown by (Lovelace and Ballas, 2013).

The same multi-level fitting approach and reference sample were used to create a synthetic population of migrants. The reference sample only contained recent migrants at this point and regional and international migrants were marked accordingly. All records in the reference sample were calibrated against target distributions of 2011 Greater Melbourne migrant population. To account for the significant part of Australia's annual population growth, migration, three different groups of migrants were generated: inter-regional, overseas temporary, and permanent overseas migrants. However, due to limited data available on migrants across different time periods, we assume that all future migrants have identical characteristics as the migrants of the base year. This is a big assumption to make but can be easily relaxed with a longitudinal dataset on migrants.

One extra step was added after the population synthesis procedure, which is to create immediate family relationships between individual whose belong in the same family. The microdata contains a variable which describes the person's relationship to a reference person of the family, or of the household, if the person does not belong in a family unit. Using that variable we were able to create parent-child and partner links for each individual in a family household. However, parent-child relationships in multi-generation households cannot be identified due to the limitation described above.

It should also be noted that, in all of our simulation runs only 1\% of the synthetic population was used, this was to significantly reduce the computing resources required for the study. The main reason why we had to reweigh the 1\% microdata was to adjust for omitting non-residents and incomplete households present in the population sample.

Empirical rates of main demographic events such as fertility, mortality, marriage and divorce can be found on ABS website. Those rates are too board geographically and have only few dimensions (usually grouped by age, sex and state). Hence, longitudinal surveys are more preferable for estimation of demographic sub-models, when other dimensions of life or a finer geographic resolution is required. Luckily, in Australia we have the Household, Income and Labour Dynamics in Australia Survey (Summerfield and Hahn, n.d.), also known as HILDA. HILDA is a household-based panel survey where many dimensions of life are captured year after year. The survey has been running for 18 consecutive waves, from 2001 to 2018. This study used the 2006 to 2016 panels to estimate its demographic sub-models. Some demographic events rarely occurred, which are also rarely captured by the survey. Due to small sample size, for those rare events, we fitted the models using pooled data, across all major Australian capital cities and the panels.

Where possible, we estimated the models separately for different groups of people based on their characteristics, such as by gender and marital status. To keep the main body of the paper concise, estimation results can be found in the Appendix section of the paper. A summary of the parameters used in the sub-models and which sub-population they applied to can be found on Table \ref{tab:sub-models}.

\hypertarget{hhsize-alignment}{%
\subsection{A household size alignment procedure}\label{hhsize-alignment}}

Household size can be difficult to simulate correctly in a dynamic microsimulation model and often not to the full extent that it can be captured validly, or aligns with an official projection. This is because there are various factors at play, from demographic to housing, and the whole decision chain does not always get modelled based on empirical data. Demographic factors such as relationship formation and dissolution (i.e.~marriage, cohabitation, divorce and break up), having children, leaving home change the size of households. These events are part of the life-cycle of families. Not only that, preferred living arrangements of overseas migrants, such as living with extended family members and renting with other people, can significantly affect household size. Moreover, housing factors such as affordability and shortage can also influence the distribution of household sizes of a population. Hence, we have devised an alignment method that allows a feasible target of household sizes to be achieved without adding more complexities to what can already found in the standard demographic components of a dynamic microsimulation model. This alignment method was applied in the immigration events to allocate new migrant families and in the divorce, break up, and leave parental home events where we replace the de facto household formation rules, discussed in the literature review section.

This method works iteratively, in each iteration it allocates one household to the household size bin that will reduce and the standard deviation of the household size differences the most at that moment. The use of standard deviation as the `scoring' function allows the errors in all of the household sizes to be balance. This process gets repeated until all the new households are assigned. An assignment can happen in two different ways. A household can either be assigned to a household size bin as a new household, or by joining an existing household with their combined household size is equal to the allocated bin. For example, let's assume a shortage of household size four and a surplus of household size two exist, if the next household to be allocated is of size two, it will be combined with an existing household of size two which immediately reduces the surplus in size two and the shortage of size four. This is an ideal outcome, however, that may not always be possible due to the unpredictable order of new households to be processed. Using a random order of new households can reduce the chance that there is a systematic bias in the process. Pseudo-code for this alignment procedure is presented in Algorithm \ref{algo:allocate} and Algorithm \ref{algo:score}.

This method is not without its drawbacks. First, it does not recognise that merging large size households can be problematic. That case would happen when large size households are to be allocated first, while big households are missing in the population. Second, this implementation of the method doesn't choose most likely households, other than their combined household size, once merged, that the new households should join. For example, one would expect people leaving parental homes for education to join group households that are also students. However, the method make no distinction between a group household and a family household of the same size. This can be easily improved by including a compatibility model, similar to a couple compatibility model, that can evaluate the compatibility of a joining household and a host household, if such data is available.

\BlankLine

\begin{algorithm}[H]
\DontPrintSemicolon
  \SetKwInOut{Input}{Inputs}
  \SetKwInOut{Output}{Output}
    \Input{$HH_{u}$, an array of unallocated households.\newline 
           $HH_{e}$, an array of existing households.\newline 
           $T$, an array of targeted household sizes, where the last index is the last category of household size (e.g. 6 or more people).}
   \BlankLine
    $B \leftarrow \textup{calculate household size bins of } HH_{e}$\;
    $D \leftarrow \textup{calculate differences between } B \text{ and } T$\;   
    \ForEach{ \textup{household} $h \in HH_e$ }
    {
        $S \leftarrow RankBestSize(h, D, T)$\;
        \ForEach{\textup{household size} $s \in S$}
        {
          \If{s == 0}{
            $h$ creates a new household\;
            $D[s] \leftarrow D[s] - 1$\;
          }
          $k \leftarrow \textup{randomly select a household of size } s \textup{ from } HH_{e}$\;
          \If{\textup{no suitable household with size} $s$} 
          {
            try the next best $s$\;
          }
          Make members of $h$ join $k$\;
          $j \leftarrow \min (\text{size of household } h + \text{size of household } k, length(T))$\;
          $D[s] = D[s] - 1$\;
          $D[j] = D[j] + 1$\;
        }
    }
\caption{Household size alignment procedure}
\label{algo:allocate}
\end{algorithm}

\BlankLine

\begin{algorithm}[H]
\DontPrintSemicolon
\SetKwInOut{Input}{Inputs}
\SetKwInOut{Output}{Output}
    \Input{
        $h$, a household. \newline
        $D$, an array that contains differences between targeted and existing household size bins. \newline
        $T$, an array that contains the target distribution of household sizes to be matched.
   }
    \Output{$R$, an array containing the ranks of the household sizes that the household, $h$, should join to minimise the mean squared of the sum of differences between $D$ and $T$, where 0 means to create a new household.}
\BlankLine
$n \leftarrow length(D)$\;
$S \leftarrow \text{an empty array with length } n$\;
$x \leftarrow \min (\text{household size of } h, n)$\;
\ForEach{$i \in 1 \textup{ to } n$ } { 
  $D^{\star} \leftarrow D$\;
  \If{i == n}
  {
    $D^{\star}[i] \leftarrow D^{\star}[i] + 1$\;
    $S[i] \leftarrow sd(D_k/T_k)$\;
    \textbf{break}\;
  }
  $j \leftarrow \min (x + i, n)$\;
  $D^{\star}[i] \leftarrow D^{\star}[i] - 1$\;
  $D^{\star}[j] \leftarrow D^{\star}[j] + 1$\;
  $S[i] \leftarrow sd(D_k/T_k)$\Comment{sd is a function to calculate stardard deviation}\;
}
$D^{\star} \leftarrow D$\Comment{Lines 14 to 16 evaluate as a new houeshold.}\;
$D^{\star}[x] \leftarrow D^{\star}[x] + 1$\;
$s \leftarrow sd(D_k/T_k)$\;
$S \leftarrow \textup{append } S \textup{ to } s$\;
$R \leftarrow \textup{sort 0 to } n \text{ from the lowest to the highest according to } S$\;
\textbf{return} $R$\;
\caption{RankBestSize}
\label{algo:score}
\end{algorithm}

\BlankLine

\hypertarget{sub-models}{%
\subsection{Sub-models}\label{sub-models}}

Our model contains 12 sub-models and within those exist sub-processes. All agents' decisions are outcomes of a Monte Carlo simulation. For example, the probability of a binary decision is determined using an appropriate model for such decision. The probability could be conditioned upon attributes of the individual making the decision. Using a pseudo random generator, a value between 0 and 1 is draw from a uniform distribution. If the randomly drawn value is less than the probability, then the individual is assumed to undergo that decision. The same technique is also used for simulating decisions with multiple outcomes. This process is also known as weighted random sampling, where the weights are the probabilities corresponding of the choices.

\hypertarget{ageing}{%
\subsubsection{Ageing}\label{ageing}}

The ageing event increases age of people by one year, since one simulation cycle of this model is equivalent to one year. This is a very important event which is also the main distinction between static and dynamic microsimulation. Ageing plays a very vital role in reflecting changes in behaviour of people such as the decision to kids or their leave parental home.

\hypertarget{birth}{%
\subsubsection{Birth}\label{birth}}

Birth is simulated in three steps. The first step is to determine the risk of giving birth for females aged 18 to 49 and simulate the risk outcomes for all, using a Monte Carlo simulation. Once the outcome of the risk is determined, the chance of giving birth to more than one baby for each female that is simulated along with gender of the newborn babies. Parent and child relationships will also be created.

\hypertarget{death}{%
\subsubsection{Death}\label{death}}

The probability of dying depends on age and sex of each individual. Once, a married person is dead, their partner will be made a widower, reflected in their marital status. Households with only children under 15 year of age left will be removed from the population. However, that rarely happened in our model so its impact to the overall result is negligible. An alternative approach would be to assign the orphans to existing households.

\hypertarget{marriage-and-cohabitation}{%
\subsubsection{Marriage and cohabitation}\label{marriage-and-cohabitation}}

New households are formed whenever individuals enter a partnership, through marriage or cohabitation. The marriage and cohabitation events are similar, procedurally. The main differences between them are in their eligibility criteria and that the marriage event is responsible for simulating two different types of marriage -- with and without premarital cohabitation -- following different procedures. Those with premarital cohabitation when they are selected to be married, it is only required that their marital status changed to `married'. In contrast, single individuals must find a suitable partner to be married or to enter a cohabitation relationship with. Hence, a mate matching market is established to match individuals who are seeking a partner. A mate matching process is required to find compatible partners. Our mate matching model assume that no one has perfect knowledge about other seekers in the market. Hence, each individual is assigned 30 potential partners to their choice set. They then have to evaluate their compatibility and the final pick is simulated using a weighted random sampling approach, where the weights are calculated based on their differences in age. This compatibility assumption can be easily changed based on the availability of empirical data that captures such information. Both partnership formation events guaranteed no `left over', by drawing the required additional number of individuals, based on their transition probabilities, to balance both of the candidate pools.

\hypertarget{divorce-and-break-up}{%
\subsubsection{Divorce and break up}\label{divorce-and-break-up}}

Relationship breakdown leads to housing stress which causes a decrease in the size of households and an increase in the number of households. When a couple is simulated to end their relationship, if they have any children, child custody will also be simulated. It is assumed that the partner that gets child custody will continues to stay together in the same household, and the other person will leave the household. If the relationship has no children involves, then the male partner has to leave the household. All individuals whose are leaving their households will either form a new lone person household or join an existing household which is up to the household size alignment algorithm, presented in Section \ref{hhsize-alignment}. For a demographic model that uses a microsimulation approach, divorce, also sometimes include de-cohabitation, is one of the ways to account for splitting of individuals into more households. Our model simulate both divorce and break up, the ending of cohabitation, exclusively of each other. This is to account for the fact that both types of partnership are different in term of relationship commitment.

\hypertarget{leaving-parental-home}{%
\subsubsection{Leaving parental home}\label{leaving-parental-home}}

There are many reasons to why people left their parental homes, for example, moving out into a couple relationship and to live with other related or non related individuals. The purpose of this event is to model the leave home decision of individuals that have never left their parental homes for other reasons that are not related to cohabitation or marriage. The cohabitation and marriage events already account for those people that will leave home to move in with their partners. Different parameters are used for males and females. In most demographic microsimulation models, leavers will form be assumed that they will go on and form one person households, this assumption undoubtedly lead to an over-representation of households in that size. Only very few models, such as Paul (2014), try to resolve this problem by introducing a roommate matching model to group leavers together to form group households. In our case, the household alignment algorithm was used to allocate leavers to households.

\hypertarget{immigration}{%
\subsubsection{Immigration}\label{immigration}}

Immigration is one of the main driver for the growth in the population of the greater Melbourne region. The number of migrants was modelled based on the ABS official migration projection. The total number of migrants expected in each year was converted into the total number of migrant households. The person to household conversion rate was calculated using the 1\% CURF data, with only the sample that migrated to the study region, for each migration type. Then that number was used to draw households from the synthetic migrant data based on their calibrated weight, that we had calibrated prior to the simulation. In each iteration, a list of migrant households was integrated with the main population using the household size alignment procedure to ensure that the overall number of households would not exceed the household projection of that year. As described in the previous section, immigration is modelled separately for each type of migrants -- permanent overseas migrants, temporary overseas migrants and interstate migrants - since they exhibit different characteristics in all levels, as observed in the 1\% CURF file.

\hypertarget{emigration}{%
\subsubsection{Emigration}\label{emigration}}

To model emigration we used the procedure presented similar to the Pageant algorithm proposed by Ch\a'enard (2000). We assumed that people emigrate as households, this is to avoid removing married people from the population, which their remaining dependent children would be marked as orphans. We randomly selected a number of individuals that aligned with the distribution of emigrants, by age and sex, from ABS. The procedure goes iteratively as the following, in each iteration an individual was randomly drawn from the population, age and sex of the individual along with its household members were checked against the target distribution. If no categories in the target distribution were in negative after removing those selected individuals characteristics, then they would be marked as emigrants and, subsequently, removed from the population. Note that, their probabilities of being selected were weighted by their household sizes. This was to make sure that households of all sizes have an equal chance of emigrating. However, this algorithm does not ensure that the simulated number of emigrants in each year will always satisfy its projection. Despite of that, we found that the differences were minute.

\hypertarget{socioeconomic-changes}{%
\subsubsection{Socioeconomic changes}\label{socioeconomic-changes}}

By making demographic decisions of individuals sensitive to changes in their socioeconomic status can make the model to be more behavioral, which can be highly desirable for many studies. In a more comprehensive model, socio-economic variables such as labour force participant can be linked to a macro-economic model and, for example, the risk of giving birth can be associated with women's labour participation status. Hence, any changes in the labour force participant rates will affect the total number of females to have children in that year. In our model, labour force status and education attainment are modelled using multinomial logistic regression models.

\hypertarget{results}{%
\section{Results}\label{results}}

\hypertarget{an-example-of-a-household-size-alignment-problem}{%
\subsection{An example of a household size alignment problem}\label{an-example-of-a-household-size-alignment-problem}}

\begin{figure}
\centering
\includegraphics{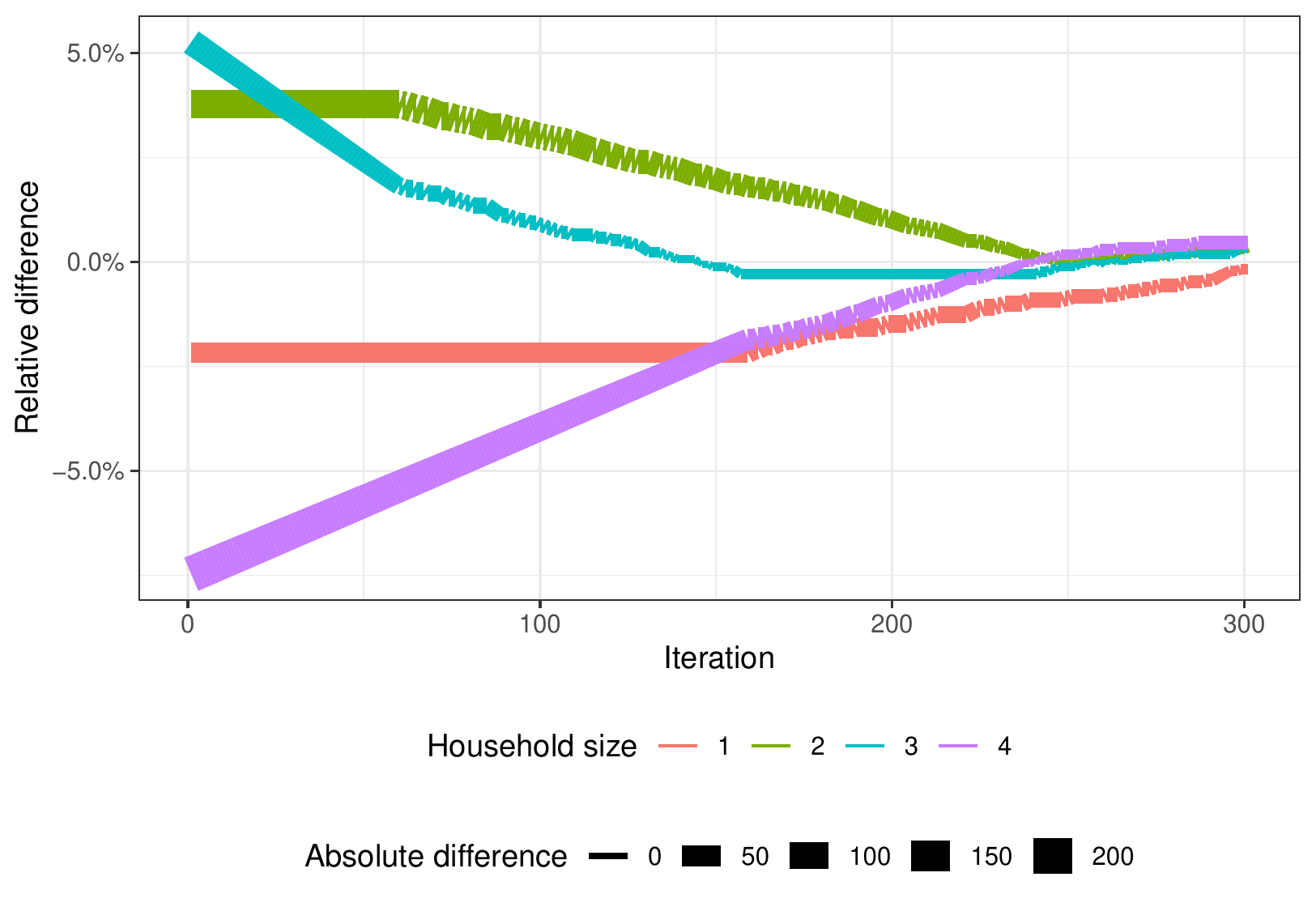}
\caption{\label{fig:household-allocation-algo-p1}Relative percentage differences between the target distribution of the existing distribution of household sizes of the example alignment problem.}
\end{figure}

This subsection illustrates how the household size alignment procedure works, using a simple example as shown on Table \ref{tab:example-alignment}. There are 100 households of each household size category, from 1 to 3, to be assigned to the existing population. However, assigning all the unallocated households as new households according to their household size would oversimulate most of the categories, except the last category, 4 or more people, since there is no new household that can be assigned to. Once we applied the household size alignment procedure to all the unallocated households, we can see very significant improvements across all the household size categories even in the 4 or more category where no unallocated household were from that size. This was because some unallocated households were allocated to combine with some existing households of size 3 then they became households of size 4 or more. Many mergers occurred between households of size 3 and new unallocated households, as can be seen on Figure \ref{fig:household-allocation-algo-p1}, specifically, from the first iteration to around 60th iteration there were a sharp decline and a surge in both of the categories, respectively. The relative percentage differences dropped to almost 0\% in all of the household sizes -- their average was around 4.5\% prior to the alignment.

\newpage

\begin{table}

\caption{\label{tab:example-alignment}An example of a household alignment problem}
\centering
\begin{tabular}[t]{lrrrr}
\toprule
\multicolumn{1}{c}{ } & \multicolumn{4}{c}{Household size} \\
\cmidrule(l{3pt}r{3pt}){2-5}
  & 1 & 2 & 3 & 4 or more\\
\midrule
(1) Unallocated households & 100 & 100 & 100 & 0\\
(2) Existing households & 2,250 & 3,300 & 1,800 & 2,600\\
(3) Target distribution & 2,300 & 3,180 & 1,710 & 2,810\\
(4) After alignment & 2,296 & 3,192 & 1,716 & 2,823\\
Relative difference before, [(2) - (3)] / (3) & -2.17 \% & 3.77 \% & 5.26 \% & -7.47 \%\\
\addlinespace
Relative difference after, [(4) - (3)] / (3) & -0.17 \% & 0.38 \% & 0.35 \% & 0.46 \%\\
\bottomrule
\end{tabular}
\end{table}

\newpage

\hypertarget{population-level}{%
\subsection{Population-level}\label{population-level}}

Table \ref{tab:population-stats-table} shows a comparison between the 1\% projected and simulated figures of population and households across five different years. For both agent types, their errors grew as the number of iterations was increasing, the same can be said about the variation in the simulated values, reported as standard deviations in the brackets. It should be pointed out that, while the total number of the starting households matched their calibration target, the total number of individuals did not. This was because the IPU algorithm could not find convergence in some categories between the 1\% microdata and the target distributions it was given. From the table above, it can be calculated that the average household size in the year 2040 would be 2.778 and 2.674 based on the simulation result and the projection, respectively. It is an indication that individual agents were forming bigger household sizes over time, in fact, greater than what implies by the projection. Since the number of immigrants added each year was already based on the immigrant projection, it is clear that births and deaths were factors in the huge discrepancy in the total number of individuals in 2040, where the model oversimulated by around 5.1\%. Unsurprisingly, all the models were estimated based on data from a historical period which only captured the then economic and social trends that influenced those events, unlike many of the assumptions made in the ABS projection that were not linear with time.

\begin{table}[!h]

\caption{\label{tab:population-stats-table}A comparison of the 1 percent simulated and observed counts of population and households.}
\centering
\begin{tabu} to \linewidth {>{\bfseries\raggedright\arraybackslash}p{1em}>{\raggedright}X>{\raggedleft}X>{\raggedleft}X>{\raggedleft}X>{\raggedleft}X>{\raggedleft}X}
\toprule
 & Measure & 2011 & 2016 & 2020 & 2030 & 2040\\
\midrule
\addlinespace[0.3em]
\multicolumn{7}{l}{\textbf{Population}}\\
\hspace{1em} & Simulated (SD) & 38,754 (0) & 44,566 (75) & 50,277 (143) & 64,101 (190) & 78,286 (229)\\

\hspace{1em} & Projected & 40,000 & 44,852 & 52,287 & 63,835 & 74,466\\

\hspace{1em} & Ratio & 0.969 & 0.994 & 0.962 & 1.004 & 1.051\\
\cmidrule{1-7}
\addlinespace[0.3em]
\multicolumn{7}{l}{\textbf{Households}}\\
\hspace{1em} & Simulated (SD) & 14,947 (0) & 16,662 (39) & 18,610 (48) & 23,289 (53) & 28,177 (56)\\

\hspace{1em} & Projected & 14,947 & 16,645 & 19,225 & 23,547 & 27,852\\

\hspace{1em} & Ratio & 1.000 & 1.001 & 0.968 & 0.989 & 1.012\\
\bottomrule
\end{tabu}
\end{table}

\hypertarget{person-level}{%
\subsection{Person-level}\label{person-level}}

Detailed historical data from the 2016 Australian Population and Housing Survey were used for benchmarking the model performance. The following comparisons are made to show that our model can reasonably capture the demographic evolution of the study region at the aggregate level. This level of comparison is generally acceptable in microsimulation studies.

Figure \ref{tab:person-level-validation-table} shows the marginal errors between the average result of 20 independent runs to their corresponding validation targets. The shares of nearly all of the simulated 5-year age groups were closely matched with their observed shares, with more than half are less than 0.20\%. The largest difference, of negative 0.51\%, can be seen in people that were in the `85 years and over' category. However, the most significant imbalances, based on their proportions, were amongst people aged between 0 to 5, 5 to 9, and 35 to 39. These categories accounted for 6.4\%, 6.2\% and 7.2\% in the observed population, respectively, which the model clearly overestimated them. While, in the three largest age categories - 20 to 34 years - were very well captured by the model, with the absolute differences of less than 0.08\% and the lowest was 0.01\%. Although the model explicitly accounted for both in- and out-migration using the administrative migration projection values, the age structure of the new migrants did not drastically change with time but only by chance of them being added, as discussed in the previous section.

It should also be noted that, the singly year age variable of the population was simulated using their five-year group to allow aging with the simulation cycle. Although we tried to ensure that the count of each singly year age group matched its observed distribution, it was not possible to simulate each individual's age jointly with their household members due to a lack of appropriate data. Some studies were able to simulate singly year age of parents conditioned on their children's, and vice versa, and also amongst each couple. This is a potential improvement to our model in a future iteration that could lessen the differences in the age structure.

Evidently, education and employment were the areas where the sub-models did not do quite well, many of the categories saw the absolute differences of well over 1.5\%, especially the share of people those were not in labour force were substantially low, compared to its observed proportion. These sub models were estimated from multiple waves of the HILDA survey with the lagged state as the independent variable. Hence, this signifies the need to revise the models with a better specification that allows the effects of other demographic variables to be accounted.

Marital status is also another area where the model was able to do well in. Out of all the categories in marital status, separation was fairly out of proportion, given its size. This disproportion was most likely the reason why the marriage category was underestimated, since only married people can go into separation.

The proportions of males and females seem to be fairly accurate. This is one of those characteristics were we were able to control. The discrepancies can only be explained by the result of the calibration of the base synthetic population and from the random drawings made to add new immigrants each year.

Overall, the results of the simulated individuals in 2016 shows that the model could reasonably simulate age, sex and marital status. However, there are still some weaknesses that should be addressed in the next iteration, such as, the specification of the multinomial logistic models used to predict the changes in education and employment status. Additionally, removing emigrants solely based on their age and sex proved to help in maintaining the total number of individuals that was close to its projection, however, it was done at the cost of creating errors in other uncontrolled categories.

\begingroup\fontsize{10}{12}\selectfont

\begin{longtabu} to \linewidth {>{\raggedright\arraybackslash}p{0.5em}>{\raggedright\arraybackslash}p{7em}>{\raggedleft}X>{\raggedleft}X>{\raggedleft}X>{\raggedleft}X}
\caption{\label{tab:person-level-validation-table}Person-level validation results in the year 2016.}\\
\toprule
 & Category & Observed & Simulated & Range (\%) & Difference\\
\midrule
\addlinespace[0.3em]
\multicolumn{6}{l}{\textbf{5-Year Age Group}}\\
\hspace{1em} & 00-04 Years & 6.4 \% & 6 \% & [5.8, 6.2] & -0.4 \%\\

\hspace{1em} & 05-09 Years & 6.2 \% & 6.6 \% & [6.5, 6.7] & 0.4 \%\\

\hspace{1em} & 10-14 Years & 5.7 \% & 5.7 \% & [5.6, 5.8] & 0 \%\\

\hspace{1em} & 15-19 Years & 6 \% & 6.3 \% & [6.2, 6.4] & 0.3 \%\\

\hspace{1em} & 20-24 Years & 7.4 \% & 7.4 \% & [7.3, 7.5] & -0.1 \%\\

\hspace{1em} & 25-29 Years & 8.1 \% & 8.1 \% & [8, 8.2] & 0 \%\\

\hspace{1em} & 30-34 Years & 8.2 \% & 8.1 \% & [7.9, 8.2] & -0.1 \%\\

\hspace{1em} & 35-39 Years & 7.3 \% & 7.7 \% & [7.6, 7.8] & 0.4 \%\\

\hspace{1em} & 40-44 Years & 7 \% & 7.3 \% & [7.2, 7.4] & 0.3 \%\\

\hspace{1em} & 45-49 Years & 6.9 \% & 7.2 \% & [7.1, 7.3] & 0.3 \%\\

\hspace{1em} & 50-54 Years & 6.2 \% & 6.3 \% & [6.3, 6.4] & 0.1 \%\\

\hspace{1em} & 55-59 Years & 5.7 \% & 5.7 \% & [5.6, 5.8] & 0 \%\\

\hspace{1em} & 60-64 Years & 4.9 \% & 4.8 \% & [4.7, 4.9] & -0.1 \%\\

\hspace{1em} & 65-69 Years & 4.4 \% & 4.2 \% & [4.1, 4.3] & -0.2 \%\\

\hspace{1em} & 70-74 Years & 3.3 \% & 3.2 \% & [3.1, 3.2] & -0.1 \%\\

\hspace{1em} & 75-79 Years & 2.5 \% & 2.3 \% & [2.3, 2.4] & -0.2 \%\\

\hspace{1em} & 80-84 Years & 1.9 \% & 1.6 \% & [1.5, 1.6] & -0.3 \%\\

\hspace{1em} & 85 Years and over & 2 \% & 1.5 \% & [1.4, 1.5] & -0.5 \%\\
\cmidrule{1-6}
\addlinespace[0.3em]
\multicolumn{6}{l}{\textbf{Education}}\\
\hspace{1em} & Advanced Diploma and Diploma Level & 8.6 \% & 8.2 \% & [8, 8.4] & -0.4 \%\\

\hspace{1em} & Bachelor Degree Level & 16.8 \% & 16.4 \% & [16.1, 16.6] & -0.5 \%\\

\hspace{1em} & Certificate Level & 11.8 \% & 14.1 \% & [13.9, 14.3] & 2.4 \%\\

\hspace{1em} & Graduate Diploma and Graduate Certificate Level & 2.3 \% & 2.5 \% & [2.4, 2.6] & 0.2 \%\\

\hspace{1em} & Not Applicable & 20.4 \% & 18.3 \% & [18.2, 18.5] & -2.1 \%\\

\hspace{1em} & Postgraduate Degree Level & 5.8 \% & 5.2 \% & [5.1, 5.3] & -0.6 \%\\

\hspace{1em} & Year 12 or Below & 34.2 \% & 35.3 \% & [35.1, 35.5] & 1.1 \%\\
\cmidrule{1-6}
\addlinespace[0.3em]
\multicolumn{6}{l}{\textbf{Employment}}\\
\hspace{1em} & Employed & 49.6 \% & 52.5 \% & [52.2, 52.8] & 3 \%\\

\hspace{1em} & Not Applicable & 19.2 \% & 18.3 \% & [18.2, 18.5] & -0.9 \%\\

\hspace{1em} & Not in the Labour Force & 27.6 \% & 23.7 \% & [23.5, 23.9] & -3.9 \%\\

\hspace{1em} & Unemployed & 3.6 \% & 5.4 \% & [5.2, 5.6] & 1.8 \%\\
\cmidrule{1-6}
\addlinespace[0.3em]
\multicolumn{6}{l}{\textbf{Marital Status}}\\
\hspace{1em} & Divorced & 6.1 \% & 6.4 \% & [6.2, 6.6] & 0.3 \%\\

\hspace{1em} & Married & 39.5 \% & 37.6 \% & [37.3, 37.9] & -1.9 \%\\

\hspace{1em} & Never Married & 29.9 \% & 32.7 \% & [32.4, 33] & 2.8 \%\\

\hspace{1em} & Not Applicable & 18.3 \% & 18.3 \% & [18.2, 18.5] & 0 \%\\

\hspace{1em} & Separated & 2.3 \% & 1.5 \% & [1.4, 1.5] & -0.9 \%\\

\hspace{1em} & Widowed & 3.9 \% & 3.5 \% & [3.4, 3.6] & -0.4 \%\\
\cmidrule{1-6}
\addlinespace[0.3em]
\multicolumn{6}{l}{\textbf{Sex}}\\
\hspace{1em} & Female & 51 \% & 51.3 \% & [51.2, 51.5] & 0.3 \%\\

\hspace{1em} & Male & 49 \% & 48.7 \% & [48.5, 48.8] & -0.3 \%\\
\bottomrule
\end{longtabu}
\endgroup{}

\hypertarget{household-level}{%
\subsection{Household-level}\label{household-level}}

\begin{figure}
\centering
\includegraphics{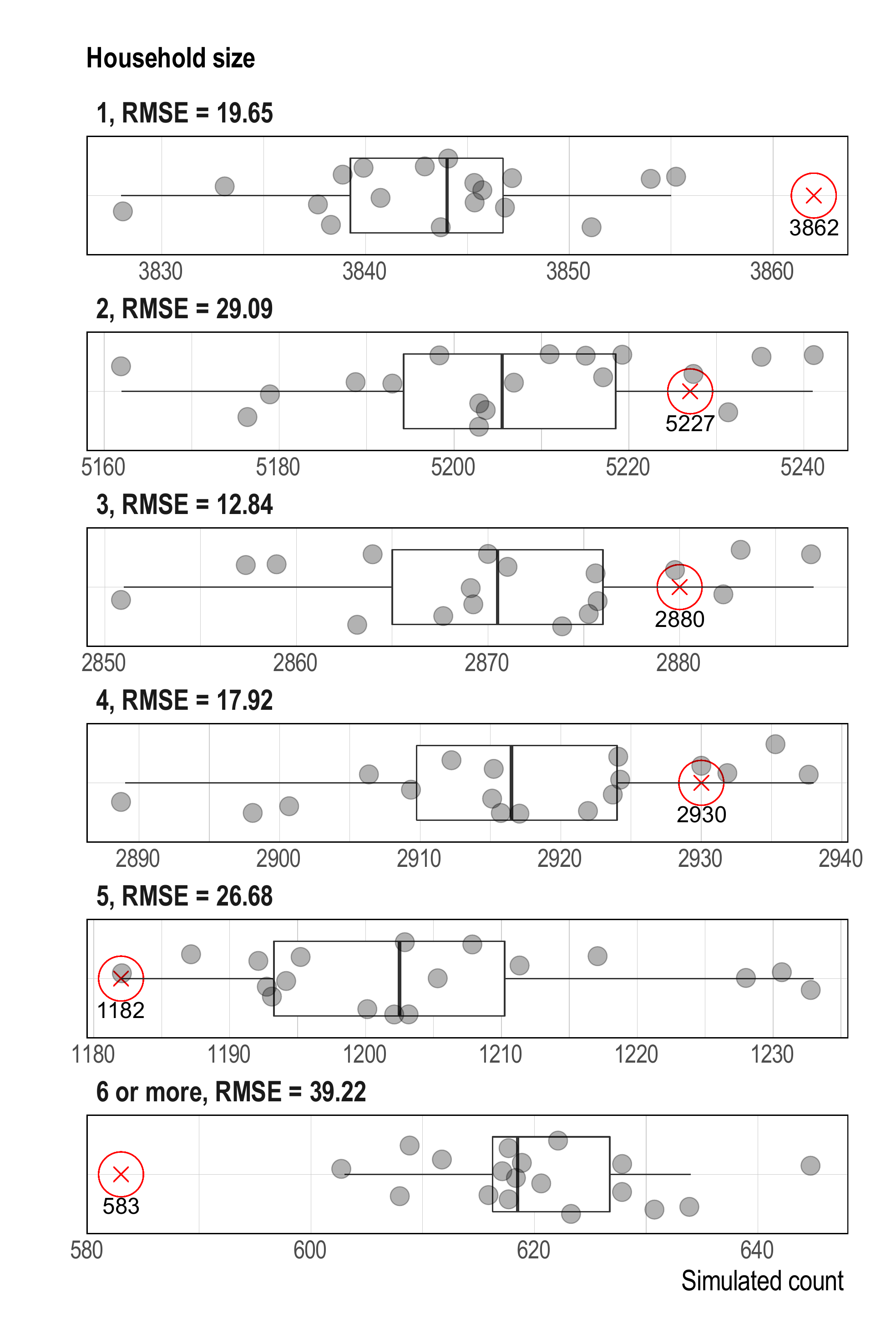}
\caption{\label{fig:hhsize-plot}Validation of household sizes in 2016. The red dots are observed counts and the grey dots are simulated counts from different runs.}
\end{figure}

\begin{figure}
\centering
\includegraphics{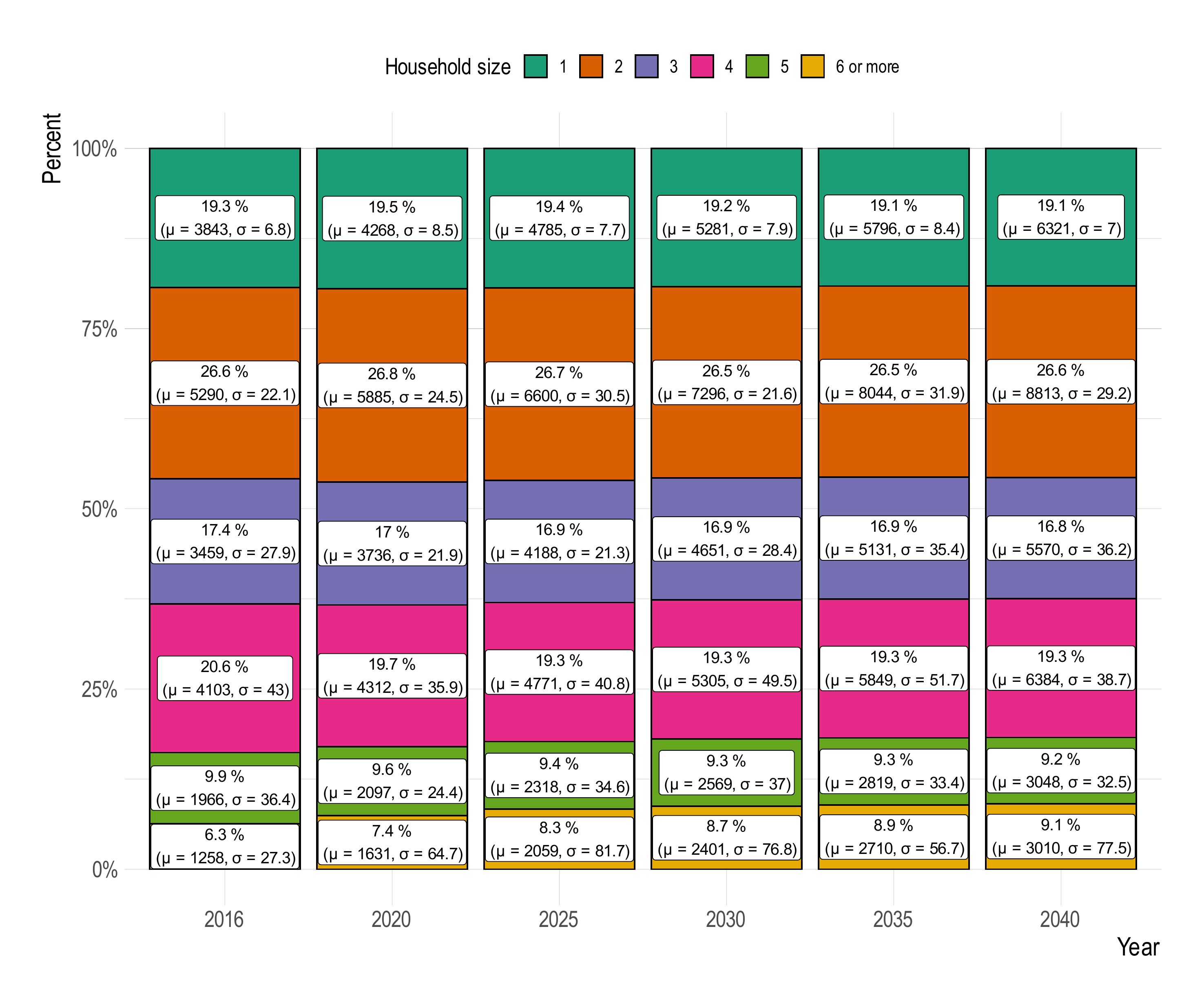}
\caption{\label{fig:household-size-projection-plot}Household size prediction vs assumption from 2016 to 2040.}
\end{figure}

Figure \ref{fig:hhsize-plot} shows that all of the simulated household sizes were closely matched with their historical targets of the same period. At first glance, it is obvious that all of the household sizes, with only an exception of one-person household, were consistently underestimated by the model. However, their RMSE values indicate that the differences were relatively small. This is evidence that the household alignment method, presented in Section \ref{hhsize-alignment}, that was used instead of the simple household formation rules worked as intended. Not only it could reasonably match with the observed numbers, the algorithm was also able to deliver the household size distribution that was consistent with our assumption that the household size distributions of all of the future periods would remain the same as in 2016, as shown in Figure \ref{fig:household-size-projection-plot}. Initially, some slight variations can be observed, then they gradually stabilised there on toward the end. In the same figure, we can see that the total number of households of size 6 or more, the smallest category, was overestimated due to merging of large households as identified in Section @ref\{hhsize-alignment\}.

\begin{figure}
\centering
\includegraphics{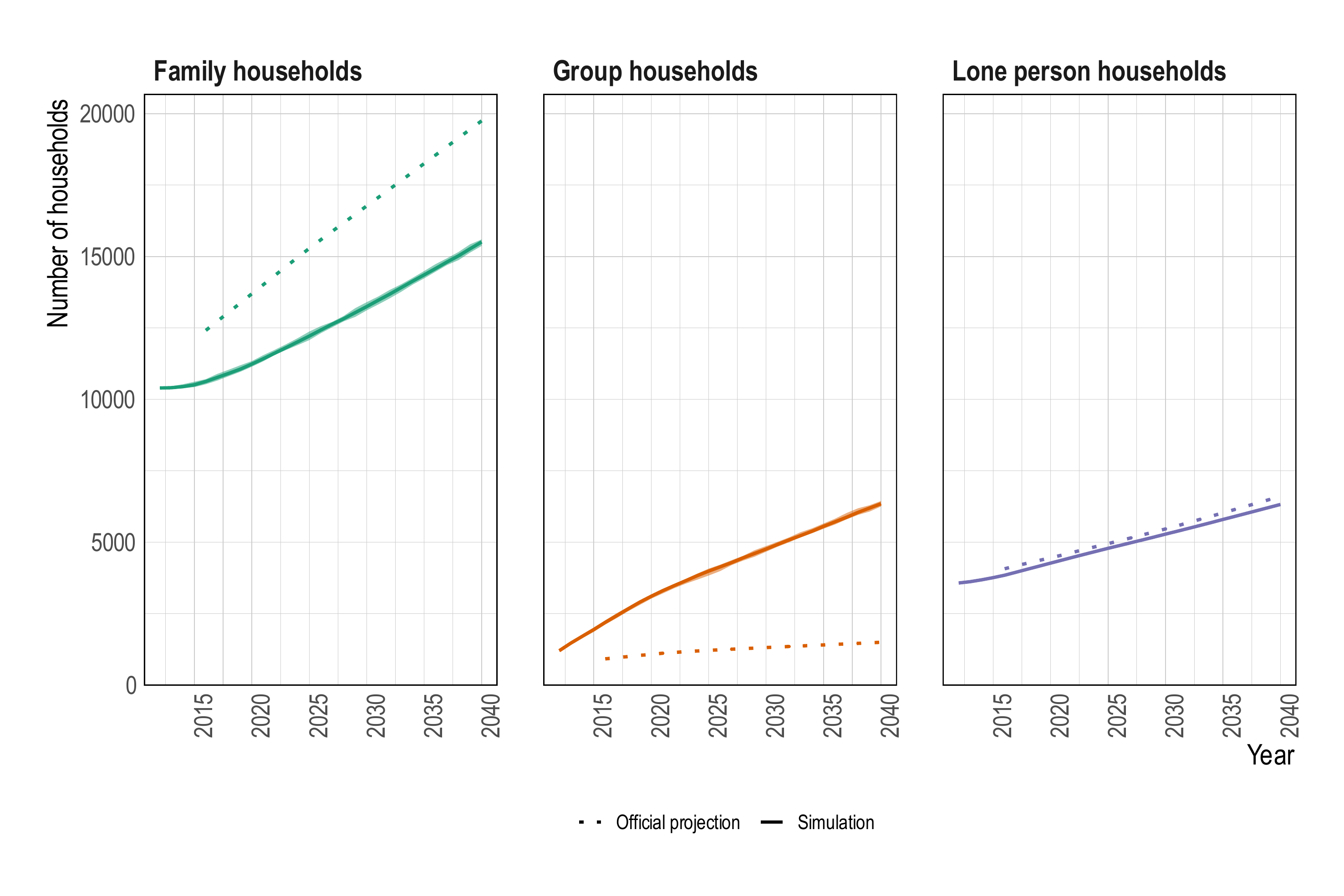}
\caption{\label{fig:household-type-projection-plot}Household type prediction vs Official projection from 2011 to 2040.}
\end{figure}

To further investigate the validity of the simulated households, we looked at how well the model could replicate three different household types by comparing the projected and simulated results, as shown in Figure \ref{fig:household-type-projection-plot}. The trends of lone person households and family households were correctly replicated, however, for the number of family households the magnitude was consistently below its projection. While the number of group households was over-represented. These discrepancies are most likely the result of the household selection criteria used, where new households only existing households to join that matched its preferred household size only, ignoring other probable compatibility metrics such as household type. Again, this was not unexpected and can be fixed with the right data that reveals migrant families' household formation behaviour.

\hypertarget{demographic-occurrences}{%
\subsection{Demographic occurrences}\label{demographic-occurences}}

\begin{figure}
\centering
\includegraphics{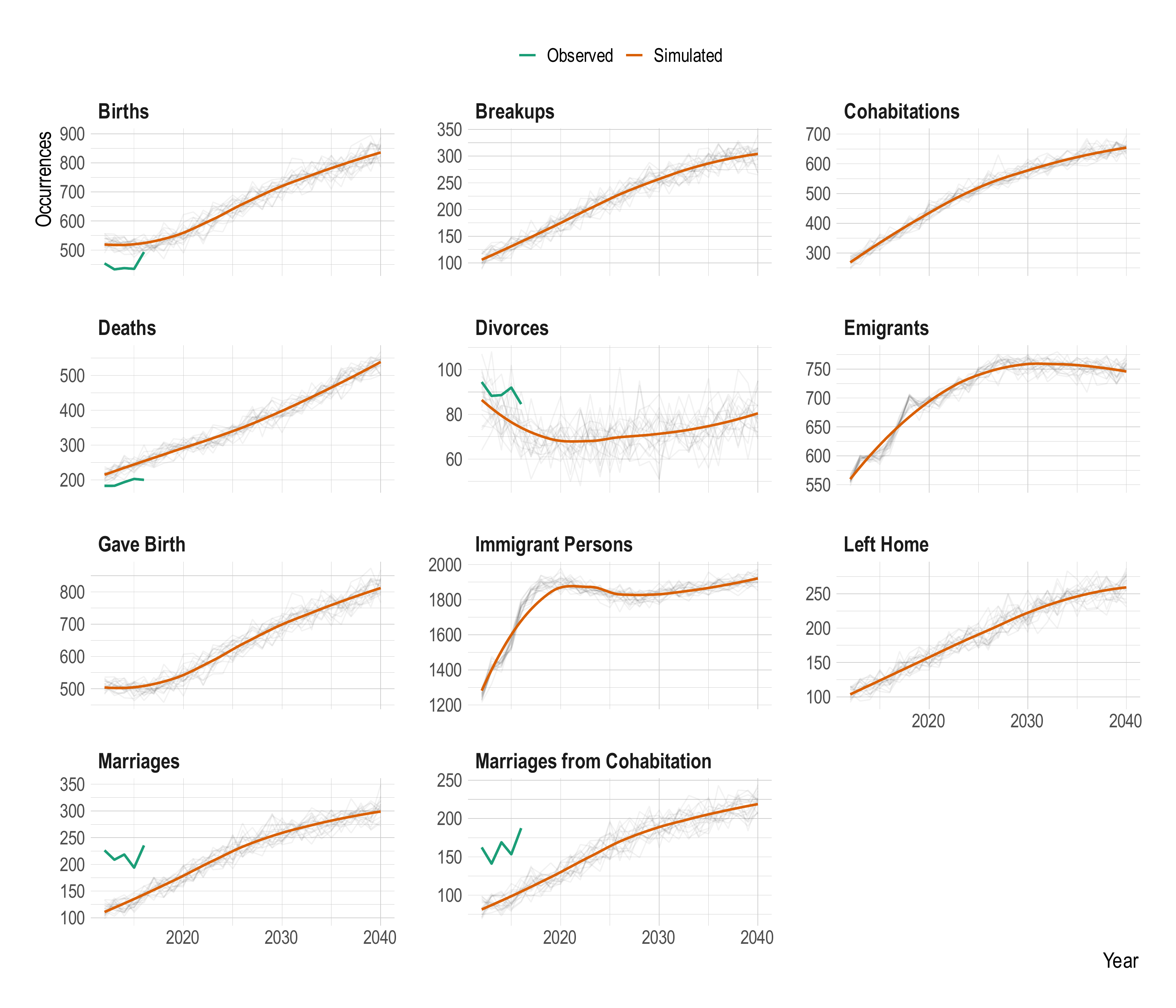}
\caption{\label{fig:demog-occurence-plot}Demographic occurrences}
\end{figure}

In addition to analysing the demographic outcomes by looking at the population's characteristics, we can also evaluate the results based on the number of occurrences of each event, as depicted in Figure \ref{fig:demog-occurence-plot}. The grey lines are from the simulation results from each simulation run, they are highly fluctuated but, overall, they all have an upward trend, which is what to be expected as the population increases. Out of all the demographic events, the number of divorces have the highest amount of variation across different periods.

Since the validation targets are for the entire State of Victoria, they had to be scaled down using a quotient calculated from their population sizes. Many demographic events included that could not be validated, due to lack of administrative data, are also included such as the number of break ups, non consensual unions, people who left parental home. As expected, the magnitudes of many demographic events are off from the approximates of their observed values.

For the migration events, a steep increase can be seen for in-migration from the beginning and levelling off to around 1900 people -- or 190,000 adjusted back from the downscaling -- from the year 2020 onward, while for out-migration a similar upward trend can be observed up to 2030 and reach a plateau with 775 of outgoing migrants per year. These trends are from the administrative projection with an assumption that Greater Melbourne will see medium interstate and overseas migration flows. However, in reality, migration is extremely volatile due to a multitude of factors -- such as immigration policies, world's economy, country's economy, global pandemic -- that are occurring in this fast-changing world.

\hypertarget{conclusion}{%
\section{Conclusion}\label{conclusion}}

In this paper we proposed a transparent and comprehensive microsimulation platform which incorporate several major elements of population evolution each of which have been discussed elsewhere but as a whole rarely come together. The proposed modular framework facilitates maintenance of the system of models as they can be easily updated. Further, a household alignment method is introduced as an alternative to the simple non data-based rules, which are deemed conservative, while they have been widely used by many demographic microsimulation models for governing the household formation and dissolution behavior of people. The proposed alignment method allows all the standard demographic events to be simulated as usual, while maintaining the distribution of household sizes of the population to be closely aligned with any pre-defined target distributions that can be variable with time. This alignment method can be applied to other demographics if a priori information is available about their distributions. We applied the method to allocate people whom experienced the relationship dissolution events (divorce and break up) or the in-migration event to the population, assuming that the proportions of household sizes were the same as what was observed in the 2016 population. It was done for three particular reasons, to lessen the number of one-person households that would otherwise be created through the conservative rule when people first leave their parental homes and those experienced relationship breakdown, and to make sure that the additions of new immigrant families wouldn't exceed a probable number of households the region was projected to have.

We showed that the model was able to reasonably project the included individual and household characteristics and the population at various periods. Some drawbacks still remain to be addressed by future studies, such as assigning people to their most likely household type, and providing a more comprehensive validation of the results. At the present, this method allows the total number of households and its household size distribution to be controlled, in an absence of appropriate empirical data and model that can capture how factors such as housing stress and demographic shifts can affect the household formation and dissolution behavior of people. Ultimately, household size has a very profound impact in many household and individual level decisions, such as in travel demand modelling, vehicle ownership decision, household expenditures, residential mobility and more. Hence, models that deal these decision levels should ensure that the household size distribution in their models should not exceed a feasible figure, such as an official household projection, to make conclusions that they draw from their results more valid.

\hypertarget{appendix}{%
\section{Appendix}\label{appendix}}

\begingroup\fontsize{8}{10}\selectfont

\begin{longtabu} to \linewidth {>{\bfseries}l>{\raggedright}X>{\raggedright}X>{\raggedright}X>{\raggedright}X>{\raggedright}X>{\raggedright}X}
\caption{\label{tab:sub-models}Summary of the sub-models}\\
\toprule
 & Sub process & Eligible to & Model & Outcomes & Covariates & Data source\\
\midrule
\endfirsthead
\caption[]{\label{tab:sub-models}Summary of the sub-models \textit{(continued)}}\\
\toprule
Sub model & Sub process & Eligible to & Model & Outcomes & Covariates & Data source\\
\midrule
\endhead
\
\endfoot
\bottomrule
\endlastfoot
\addlinespace[0.3em]
\multicolumn{7}{l}{\textbf{Birth}}\\
\hspace{1em} & Fertility & Females aged 16 and above & Logistic regression & Binary & Age, parity, age of youngest child & HILDA waves 2006 - 2016\\
\cmidrule{2-7}
\hspace{1em} & Multiplicity & Females giving birth & Rate-based & Categorical &  & ABS Statistic\\
\cmidrule{2-7}
\hspace{1em} & Sex of newborn & Newborns & Rate-based & Categorical &  & ABS Statistic\\
\cmidrule{1-7}
\addlinespace[0.3em]
\multicolumn{7}{l}{\textbf{Death}}\\
\hspace{1em} & Dying & All individuals & Logistic regression & Binary & Age and sex & ABS Statistic\\
\cmidrule{1-7}
\addlinespace[0.3em]
\multicolumn{7}{l}{\textbf{Marriage}}\\
\hspace{1em} & Direct marriage & Cohabiting males & Logistic regression & Binary & Age, marital status & HILDA waves 2006 - 2016\\
\cmidrule{2-7}
\hspace{1em} & Indirect marriage & Never married females aged above 18 & Logistic regression & Binary & Age, marital status & HILDA waves 2006 - 2016\\
\cmidrule{3-7}
\hspace{1em}\hspace{1em} &  & Never married males aged above 18 & Logistic regression & Binary & Age, marital status & HILDA waves 2006 - 2016\\
\cmidrule{3-7}
\hspace{1em}\hspace{1em} &  & Previosly married females aged above 18 & Logistic regression & Binary & Age, marital status & HILDA waves 2006 - 2016\\
\cmidrule{3-7}
\hspace{1em}\hspace{1em} &  & Previosly married males aged above 18 & Logistic regression & Binary & Age, marital status & HILDA waves 2006 - 2016\\
\cmidrule{2-7}
\hspace{1em}\hspace{1em} & Partner score & Individuals seeking partner & Exponential function & Numeric & Age difference & \\
\cmidrule{1-7}
\addlinespace[0.3em]
\multicolumn{7}{l}{\textbf{Cohabitation}}\\
\hspace{1em} & Indirect cohabitation & Never married females aged above 18 & Logistic regression & Binary & Age, marital status & HILDA waves 2006 - 2016\\
\cmidrule{3-7}
 &  & Never married males aged above 18 & Logistic regression & Binary & Age, marital status & HILDA waves 2006 - 2016\\
\cmidrule{3-7}
 &  & Previosly married females aged above 18 & Logistic regression & Binary & Age, marital status & HILDA waves 2006 - 2016\\
\cmidrule{3-7}
 &  & Previosly married males aged above 18 & Logistic regression & Binary & Age, marital status & HILDA waves 2006 - 2016\\
\cmidrule{2-7}
 & Partner score & Individuals seeking partner & Exponential function & Numeric & Age difference & \\
\cmidrule{1-7}
\addlinespace[0.3em]
\multicolumn{7}{l}{\textbf{Breakup}}\\
\hspace{1em} & Breakup decision & Never married females aged above 18 & Logistic regression & Binary & Age, marital status & HILDA waves 2006 - 2016\\
\cmidrule{2-7}
\hspace{1em} & Moveout decision & Never married males aged above 18 & Logistic regression & Binary & Age, marital status & HILDA waves 2006 - 2016\\
\cmidrule{1-7}
\addlinespace[0.3em]
\multicolumn{7}{l}{\textbf{Divorce}}\\
\hspace{1em} & Decision to divorce & Previously married females aged above 18 & Logistic regression & Binary & Age, marital status & HILDA waves 2006 - 2016\\
\cmidrule{2-7}
\hspace{1em} & Moveout decision & Previously married males aged above 18 & Logistic regression & Binary & Age, marital status & HILDA waves 2006 - 2016\\
\cmidrule{1-7}
\addlinespace[0.3em]
\multicolumn{7}{l}{\textbf{Leavehome}}\\
\hspace{1em} & Decision to leave parental home & Female children aged between 18 to 40 & Logistic regression & Binary & Age & HILDA waves 2006 - 2016\\
\cmidrule{1-7}
\addlinespace[0.3em]
\multicolumn{7}{l}{\textbf{Socioeconomics}}\\
\hspace{1em} & Education & All individuals aged above 15 & Multinomai regression & Categorical & Education & HILDA waves 2006 - 2016\\
\cmidrule{2-7}
\hspace{1em} & Employment & All individuals aged above 15 & Multinomai regression & Categorical & Employment & HILDA waves 2006 - 2016\\
\cmidrule{1-7}
\addlinespace[0.3em]
\multicolumn{7}{l}{\textbf{Emigration}}\\
\hspace{1em} & Overseas migrants & All individuals & Algorithmic & Binary & 5-year age, sex & ABS migration projection\\
\cmidrule{1-7}
\addlinespace[0.3em]
\multicolumn{7}{l}{\textbf{Immigration}}\\
\hspace{1em} & Overseas temoporary migrants &  & Weighted draw & Numeric & Calibrated weights & 2011 CURF and ABS migration projection\\
\cmidrule{2-7}
\hspace{1em} & Overseas permanent migrants &  & Weighted draw & Numeric & Calibrated weights & 2011 CURF and ABS migration projection\\
\cmidrule{2-7}
\hspace{1em} & Inter-regional migrants &  & Weighted draw & Numeric & Calibrated weights & 2011 CURF and ABS migration projection\\*
\end{longtabu}
\endgroup{}

\begin{table}[H]

\caption{\label{tab:model-single-fertility}Single females' fertility model}
\centering
\begin{tabular}[t]{lrrrr}
\toprule
Term & Estimate & Std.error & Statistic & P.value\\
\midrule
Intercept & -11.853 & 5.230 & -2.266 & 0.023\\
Age & 0.610 & 0.374 & 1.632 & 0.103\\
Age\textasciicircum{}2 & -0.012 & 0.006 & -1.866 & 0.062\\
Employed & -1.036 & 0.631 & -1.642 & 0.101\\
Has one child & 2.607 & 0.846 & 3.081 & 0.002\\
\addlinespace
Has two or more children & 1.946 & 0.998 & 1.951 & 0.051\\
\bottomrule
\end{tabular}
\end{table}

\begin{table}[H]

\caption{\label{tab:model-cohabiting-fertility}Cohabiting females' fertility model}
\centering
\begin{tabular}[t]{lrrrr}
\toprule
Term & Estimate & Std.error & Statistic & P.value\\
\midrule
Intercept & -11.089 & 5.396 & -2.055 & 0.040\\
Age & 0.576 & 0.333 & 1.728 & 0.084\\
Age\textasciicircum{}2 & -0.008 & 0.005 & -1.655 & 0.098\\
Age of youngest child & -0.054 & 0.093 & -0.586 & 0.558\\
Employed & -1.672 & 0.460 & -3.638 & 0.000\\
\bottomrule
\end{tabular}
\end{table}

\begin{table}[H]

\caption{\label{tab:model-married-fertility}Married females' fertility model}
\centering
\begin{tabular}[t]{lrrrr}
\toprule
Term & Estimate & Std.error & Statistic & P.value\\
\midrule
Intercept & -5.225 & 1.375 & -3.801 & 0.000\\
Age of youngest child & -0.073 & 0.043 & -1.700 & 0.089\\
Employed & -0.340 & 0.216 & -1.580 & 0.114\\
Age: 15 - 19 years with no child & -8.039 & 520.429 & -0.015 & 0.988\\
Age: 20 - 25 years with no child & 2.930 & 1.590 & 1.843 & 0.065\\
\addlinespace
Age: 25 - 29 years with no child & 4.112 & 1.397 & 2.943 & 0.003\\
Age: 30 - 34 years with no child & 4.405 & 1.392 & 3.164 & 0.002\\
Age: 35 - 39 years with no child & 4.944 & 1.408 & 3.511 & 0.000\\
Age: 40 - 44 years with no child & 2.985 & 1.500 & 1.989 & 0.047\\
Age: 45 - 49 years with no child & 2.325 & 1.755 & 1.324 & 0.185\\
\addlinespace
Age: 20 - 25 years with one child & 18.721 & 620.648 & 0.030 & 0.976\\
Age: 25 - 29 years with one child & 5.117 & 1.383 & 3.700 & 0.000\\
Age: 30 - 34 years with one child & 4.425 & 1.370 & 3.229 & 0.001\\
Age: 35 - 39 years with one child & 4.409 & 1.369 & 3.220 & 0.001\\
Age: 40 - 44 years with one child & 3.775 & 1.378 & 2.740 & 0.006\\
\addlinespace
Age: 45 - 49 years with one child & 1.844 & 1.832 & 1.006 & 0.314\\
Age: 20 - 25 years with two or more children & 3.747 & 1.468 & 2.553 & 0.011\\
Age: 25 - 29 years with two or more children & 4.031 & 1.370 & 2.942 & 0.003\\
Age: 30 - 34 years with two or more children & 2.866 & 1.365 & 2.099 & 0.036\\
Age: 35 - 39 years with two or more children & 0.890 & 1.471 & 0.605 & 0.545\\
\bottomrule
\end{tabular}
\end{table}

\begin{table}[H]

\caption{\label{tab:model-cohab-male-never-married}Never married males' cohabitation model}
\centering
\begin{tabular}[t]{lrrrr}
\toprule
Term & Estimate & Std.error & Statistic & P.value\\
\midrule
Intercept & -13.159 & 1.929 & -6.820 & 0.00\\
Age & 0.687 & 0.133 & 5.149 & 0.00\\
Age\textasciicircum{}2 & -0.011 & 0.002 & -4.988 & 0.00\\
Employed & 0.729 & 0.283 & 2.577 & 0.01\\
\bottomrule
\end{tabular}
\end{table}

\begin{table}[H]

\caption{\label{tab:model-cohab-male-married-before}Priorly married males' cohabitation model}
\centering
\begin{tabular}[t]{lrrrr}
\toprule
Term & Estimate & Std.error & Statistic & P.value\\
\midrule
Intercept & -12.525 & 5.030 & -2.490 & 0.013\\
Age & 0.406 & 0.240 & 1.688 & 0.091\\
Age\textasciicircum{}2 & -0.005 & 0.003 & -1.840 & 0.066\\
Employed & 0.539 & 0.811 & 0.665 & 0.506\\
\bottomrule
\end{tabular}
\end{table}

\begin{table}[H]

\caption{\label{tab:model-cohab-female-never-married}Never married females' cohabitation model}
\centering
\begin{tabular}[t]{lrrrr}
\toprule
Term & Estimate & Std.error & Statistic & P.value\\
\midrule
Intercept & -10.406 & 1.740 & -5.982 & 0.000\\
Age & 0.548 & 0.125 & 4.396 & 0.000\\
Age\textasciicircum{}2 & -0.009 & 0.002 & -4.253 & 0.000\\
Employed & 0.151 & 0.263 & 0.572 & 0.567\\
\bottomrule
\end{tabular}
\end{table}

\begin{table}[H]

\caption{\label{tab:model-cohab-female-married-before}Priorly married females' cohabitation model}
\centering
\begin{tabular}[t]{lrrrr}
\toprule
Term & Estimate & Std.error & Statistic & P.value\\
\midrule
Intercept & -11.440 & 4.684 & -2.442 & 0.015\\
Age & 0.368 & 0.225 & 1.640 & 0.101\\
Age\textasciicircum{}2 & -0.004 & 0.003 & -1.696 & 0.090\\
Employed & -1.025 & 0.382 & -2.683 & 0.007\\
\bottomrule
\end{tabular}
\end{table}

\begin{table}[H]

\caption{\label{tab:model-marriage-male-never-married}Never married males' direct marriage model}
\centering
\begin{tabular}[t]{lrrrr}
\toprule
Term & Estimate & Std.error & Statistic & P.value\\
\midrule
Intercept & -26.424 & 9.636 & -2.742 & 0.006\\
Age & 1.354 & 0.622 & 2.176 & 0.030\\
Age\textasciicircum{}2 & -0.021 & 0.010 & -2.076 & 0.038\\
\bottomrule
\end{tabular}
\end{table}

\begin{table}[H]

\caption{\label{tab:model-marriage-male-married-before}Priorly married males' direct marriage model}
\centering
\begin{tabular}[t]{lrrrr}
\toprule
Term & Estimate & Std.error & Statistic & P.value\\
\midrule
Intercept & -14810.575 & 307025.01 & -0.048 & 0.962\\
Age & 638.060 & 13206.92 & 0.048 & 0.961\\
Age\textasciicircum{}2 & -6.873 & 142.01 & -0.048 & 0.961\\
\bottomrule
\end{tabular}
\end{table}

\begin{table}[H]

\caption{\label{tab:model-marriage-female-never-married}Never married females' direct marriage model}
\centering
\begin{tabular}[t]{lrrrr}
\toprule
Term & Estimate & Std.error & Statistic & P.value\\
\midrule
Intercept & -16.509 & 5.399 & -3.058 & 0.002\\
Age & 0.769 & 0.365 & 2.106 & 0.035\\
Age\textasciicircum{}2 & -0.012 & 0.006 & -1.958 & 0.050\\
\bottomrule
\end{tabular}
\end{table}

\begin{table}[H]

\caption{\label{tab:model-marriage-female-married-before}Priorly married females' direct marriage model}
\centering
\begin{tabular}[t]{lrrrr}
\toprule
Term & Estimate & Std.error & Statistic & P.value\\
\midrule
Intercept & -12.718 & 19.571 & -0.650 & 0.516\\
Age & 0.163 & 0.850 & 0.192 & 0.848\\
Age\textasciicircum{}2 & -0.001 & 0.009 & -0.098 & 0.922\\
\bottomrule
\end{tabular}
\end{table}

\begin{table}[H]

\caption{\label{tab:model-breakup-male}Breakup model for males}
\centering
\begin{tabular}[t]{lrrrr}
\toprule
Term & Estimate & Std.error & Statistic & P.value\\
\midrule
Intercept & -0.049 & 0.528 & -0.093 & 0.926\\
Age\textasciicircum{}2 & -0.001 & 0.000 & -2.413 & 0.016\\
Holds a degree & -0.145 & 0.435 & -0.333 & 0.739\\
Employed & -1.906 & 0.424 & -4.492 & 0.000\\
\bottomrule
\end{tabular}
\end{table}

\begin{table}[H]

\caption{\label{tab:model-breakup-female}Breakup model for females}
\centering
\begin{tabular}[t]{lrrrr}
\toprule
Term & Estimate & Std.error & Statistic & P.value\\
\midrule
Intercept & -0.434 & 0.402 & -1.078 & 0.281\\
Age\textasciicircum{}2 & -0.001 & 0.000 & -3.808 & 0.000\\
Holds a degree & -0.842 & 0.394 & -2.135 & 0.033\\
Employed & -0.381 & 0.337 & -1.133 & 0.257\\
\bottomrule
\end{tabular}
\end{table}

\begin{table}[H]

\caption{\label{tab:model-divorce-male}Divorce model for males}
\centering
\begin{tabular}[t]{lrrrr}
\toprule
Term & Estimate & Std.error & Statistic & P.value\\
\midrule
Intercept & -1.691 & 0.710 & -2.382 & 0.017\\
Age & -0.037 & 0.018 & -2.026 & 0.043\\
Has children & -0.694 & 0.378 & -1.837 & 0.066\\
Holds a degree & -1.494 & 0.498 & -3.002 & 0.003\\
\bottomrule
\end{tabular}
\end{table}

\begin{table}[H]

\caption{\label{tab:model-divorce-female}Divorce model for females}
\centering
\begin{tabular}[t]{lrrrr}
\toprule
Term & Estimate & Std.error & Statistic & P.value\\
\midrule
Intercept & -2.062 & 0.760 & -2.712 & 0.007\\
Age & -0.052 & 0.019 & -2.707 & 0.007\\
Has children & -0.164 & 0.443 & -0.370 & 0.712\\
Holds a degree & -0.067 & 0.363 & -0.184 & 0.854\\
\bottomrule
\end{tabular}
\end{table}

\newpage
\begin{landscape}

\end{landscape}
\newpage

\clearpage

\clearpage

\hypertarget{references}{%
\section*{References}\label{references}}
\addcontentsline{toc}{section}{References}

\hypertarget{refs}{}
\begin{cslreferences}
\leavevmode\hypertarget{ref-anderson1997models}{}%
Anderson, J., 1997. Models for retirement policy analysis, report to the society of actuaries. Schaumburg, Illinois Society of Actuaries (SOA).

\leavevmode\hypertarget{ref-bacon2007appsim}{}%
Bacon, B., Pennec, S., 2007. APPSIM-Modelling family formation and dissolution. National Centre for Social and Economic Modelling.

\leavevmode\hypertarget{ref-baekgaardMicromacroLinkageAlignment2002}{}%
Baekgaard, H., 2002. Micro-macro linkage and the alignment of transition processes. Technical paper.

\leavevmode\hypertarget{ref-ballasSimBritainSpatialMicrosimulation2005}{}%
Ballas, D., Clarke, G., Dorling, D., Eyre, H., Thomas, B., Rossiter, D., 2005. SimBritain: A spatial microsimulation approach to population dynamics. Population, Space and Place 11, 13--34. doi:\href{https://doi.org/10.1002/psp.351}{10.1002/psp.351}

\leavevmode\hypertarget{ref-BenensonGeosimulationautomatabasedmodeling2004}{}%
Benenson, I., Torrens, P.M., 2004. Geosimulation: Automata-based modeling of urban phenomena. John Wiley \& Sons, Hoboken, NJ.

\leavevmode\hypertarget{ref-BodenmannModellingfirmre2011a}{}%
Bodenmann, B.R., 2011. Modelling firm (re-)location choice in UrbanSim, in:. Louvain-la-Neuve: European Regional Science Association (ERSA).

\leavevmode\hypertarget{ref-chenard2000individual}{}%
Ch\a'enard, D., 2000. Individual alignment and group processing: An application to migration processes in DYNACAN. OCCASIONAL PAPERS-UNIVERSITY OF CAMBRIDGE DEPARTMENT OF APPLIED ECONOMICS 238--250.

\leavevmode\hypertarget{ref-chingcuancoILUTEDemographicMicrosimulation2018}{}%
Chingcuanco, F., Miller, E.J., 2018. The ILUTE Demographic Microsimulation Model for the Greater Toronto-Hamilton Area: Current Operational Status and Historical Validation, in: Thill, J.-C., Dragicevic, S. (Eds.), GeoComputational Analysis and Modeling of Regional Systems. Springer International Publishing, Cham, pp. 167--187. doi:\href{https://doi.org/10.1007/978-3-319-59511-5_10}{10.1007/978-3-319-59511-5\_10}

\leavevmode\hypertarget{ref-clarke1986demographic}{}%
Clarke, M., 1986. Demographic processes and household dynamics: A microsimulation approach.

\leavevmode\hypertarget{ref-dekkersModellingImmigrationEmigration2015}{}%
Dekkers, G., 2015. On the modelling of immigration and emigration using LIAM2. doi:\href{https://doi.org/10.13140/rg.2.1.4373.8967}{10.13140/rg.2.1.4373.8967}

\leavevmode\hypertarget{ref-dekkers2010consequences}{}%
Dekkers, G., Buslei, H., Cozzolino, M., Desmet, R., Geyer, J., Hofmann, D., Raitano, M., Steiner, V., Tanda, P., Tedeschi, S., others, 2010. What are the consequences of the european awg-projections on the adequacy of pensions? An application of the dynamic micro simulation model midas for belgium, germany and italy.

\leavevmode\hypertarget{ref-deloitteHousingAspirationsNew2011}{}%
Deloitte, 2011. The housing aspirations of new settlers to Australia.

\leavevmode\hypertarget{ref-dumontRightOrderingSequence2018}{}%
Dumont, M., Barthelemy, J., Huynh, N., Carletti, T., 2018. Towards the Right Ordering of the Sequence of Models for the Evolution of a Population Using Agent-Based Simulation. Journal of Artificial Societies and Social Simulation 21. doi:\href{https://doi.org/10.18564/jasss.3790}{10.18564/jasss.3790}

\leavevmode\hypertarget{ref-eluruPopulationUpdatingSystem2008}{}%
Eluru, N., Pinjari, A., Guo, J., Sener, I., Srinivasan, S., Copperman, R., Bhat, C., 2008. Population Updating System Structures and Models Embedded in the Comprehensive Econometric Microsimulator for Urban Systems. Transportation Research Record: Journal of the Transportation Research Board 2076, 171--182. doi:\href{https://doi.org/10.3141/2076-19}{10.3141/2076-19}

\leavevmode\hypertarget{ref-fatmiMicrosimulationVehicleTransactions2018}{}%
Fatmi, M.R., Habib, M.A., 2018. Microsimulation of vehicle transactions within a life-oriented integrated urban modeling system. Transportation Research Part A: Policy and Practice 116, 497--512. doi:\href{https://doi.org/10.1016/j.tra.2018.06.029}{10.1016/j.tra.2018.06.029}

\leavevmode\hypertarget{ref-figariMicrosimulationPolicyAnalysis2015}{}%
Figari, F., Paulus, A., Sutherland, H., 2015. Microsimulation and Policy Analysis, in: Handbook of Income Distribution. Elsevier, pp. 2141--2221. doi:\href{https://doi.org/10.1016/B978-0-444-59429-7.00025-X}{10.1016/B978-0-444-59429-7.00025-X}

\leavevmode\hypertarget{ref-flood2005sesim}{}%
Flood, L., Jansson, F., Pettersson, T., Pettersson, T., Sundberg, O., Westerberg, A., 2005. SESIM iii--a swedish dynamic microsimulation model. Handbook of sesim, Ministry of Finance, Stocholm.

\leavevmode\hypertarget{ref-fredriksen1998projections}{}%
Fredriksen, D., 1998. Projections of population, education, labour supply and public pension benefits. Analyses with the dynamic microsimulation model MOSART. Statistisk sentralbyrå.

\leavevmode\hypertarget{ref-fukawaHouseholdProjectionIts2011}{}%
Fukawa, T., 2011. Household Projection and Its Application to Health/Long-Term Care Expenditures in Japan Using INAHSIM-II. Social Science Computer Review 29, 52--66. doi:\href{https://doi.org/10.1177/0894439310370097}{10.1177/0894439310370097}

\leavevmode\hypertarget{ref-gallerMicrosimulationHouseholdFormation1988}{}%
Galler, H., 1988. Microsimulation of household formation and dissolution 12.

\leavevmode\hypertarget{ref-geardSyntheticPopulationDynamics2013}{}%
Geard, N., McCaw, J.M., Dorin, A., Korb, K.B., McVernon, J., 2013. Synthetic Population Dynamics: A Model of Household Demography. Journal of Artificial Societies and Social Simulation 16. doi:\href{https://doi.org/10.18564/jasss.2098}{10.18564/jasss.2098}

\leavevmode\hypertarget{ref-gouuas1992travel}{}%
Goulias, K., Kitamura, R., 1992. Travel demand forecasting with dynamic microsimulation. Transportation Research Record 1357, 8--17.

\leavevmode\hypertarget{ref-hardingAPPSIMAustralianDynamic2007}{}%
Harding, A., 2007. APPSIM: The Australian Dynamic Population and Policy Microsimulation Model 11.

\leavevmode\hypertarget{ref-harmonOverviewLabourMarket2018}{}%
Harmon, A., Miller, E.J., 2018. Overview of a labour market microsimulation model. Procedia Computer Science, The 9th International Conference on Ambient Systems, Networks and Technologies (ANT 2018) / The 8th International Conference on Sustainable Energy Information Technology (SEIT-2018) / Affiliated Workshops 130, 172--179. doi:\href{https://doi.org/10.1016/j.procs.2018.04.027}{10.1016/j.procs.2018.04.027}

\leavevmode\hypertarget{ref-inagakiDynamicMicrosimulationModel2018}{}%
Inagaki, S., 2018. Dynamic Microsimulation Model of Impoverishment Among Elderly Women in Japan. Frontiers in Physics 6. doi:\href{https://doi.org/10.3389/fphy.2018.00022}{10.3389/fphy.2018.00022}

\leavevmode\hypertarget{ref-lawsonMethodsToolsMicrosimulation2014}{}%
Lawson, T., 2014. Methods and Tools for the Microsimulation of Household Expenditure (PhD thesis).

\leavevmode\hypertarget{ref-lawsonMethodsToolsMicrosimulation2014a}{}%
Lawson, T., Anderson, B., 2014. Methods and Tools for the Microsimulation of Household Expenditure (PhD Thesis). University of Essex.

\leavevmode\hypertarget{ref-liSurveyDynamicMicrosimulation2013}{}%
Li, J., O'Donoghue, C., 2013. A survey of dynamic microsimulation models: Uses, model structure and methodology. INTERNATIONAL JOURNAL OF MICROSIMULATION 6, 3--55.

\leavevmode\hypertarget{ref-liEvaluatingBinaryAlignment2014}{}%
Li, J., O'Donoghue, C., 2014. Evaluating Binary Alignment Methods in Microsimulation Models. Journal of Artificial Societies and Social Simulation 17. doi:\href{https://doi.org/10.18564/jasss.2334}{10.18564/jasss.2334}

\leavevmode\hypertarget{ref-lovelaceTruncateReplicateSample2013}{}%
Lovelace, R., Ballas, D., 2013. ``Truncate, replicate, sample'': A method for creating integer weights for spatial microsimulation. Computers, Environment and Urban Systems 41, 1--11. doi:\href{https://doi.org/10.1016/j.compenvurbsys.2013.03.004}{10.1016/j.compenvurbsys.2013.03.004}

\leavevmode\hypertarget{ref-mckay2003dynamic}{}%
McKay, S., 2003. Dynamic microsimulation at the us social security administration, in: International Microsimulation Conference on Population Ageing and Health, Canberra.

\leavevmode\hypertarget{ref-millerCaseMicrosimulationFrameworks2018}{}%
Miller, E.J., 2018. The case for microsimulation frameworks for integrated urban models. Journal of Transport and Land Use 11. doi:\href{https://doi.org/10.5198/jtlu.2018.1257}{10.5198/jtlu.2018.1257}

\leavevmode\hypertarget{ref-moeckelConstraintsHouseholdRelocation2016}{}%
Moeckel, R., 2016. Constraints in household relocation: Modeling land-use/transport interactions that respect time and monetary budgets. Journal of Transport and Land Use 10. doi:\href{https://doi.org/10.5198/jtlu.2015.810}{10.5198/jtlu.2015.810}

\leavevmode\hypertarget{ref-morandDemographicModellingState2010}{}%
Morand, E., Toulemon, L., Pennec, S., Baggio, R., Billari, F., 2010. Demographic modelling: The state of the art. Paris, INED. FP7-244557 Projet SustainCity 39.

\leavevmode\hypertarget{ref-mullerGeneralizedApproachPopulation2017}{}%
Müller, K., 2017. A Generalized Approach to Population Synthesis. ETH Zurich. doi:\href{https://doi.org/10.3929/ethz-b-000171586}{10.3929/ethz-b-000171586}

\leavevmode\hypertarget{ref-o2009life}{}%
O'Donoghue, C., Lennon, J., Hynes, S., others, 2009. The Life-cycle Income Analysis Model (LIAM): A study of a flexible dynamic microsimulation modelling computing framework. International Journal of Microsimulation 2, 16--31.

\leavevmode\hypertarget{ref-orcuttNewTypeSocioEconomic1957}{}%
Orcutt, G.H., 1957. A New Type of Socio-Economic System. The Review of Economics and Statistics 39, 116--123. doi:\href{https://doi.org/10.2307/1928528}{10.2307/1928528}

\leavevmode\hypertarget{ref-paulActivityBasedTravelBehavior2014}{}%
Paul, S., 2014. Activity-Based Travel Behavior Analysis and Demand Forecasting (PhD thesis). ARIZONA STATE UNIVERSITY.

\leavevmode\hypertarget{ref-RashidiDynamicHousingSearch2015}{}%
Rashidi, T.H., 2015. Dynamic Housing Search Model Incorporating Income Changes, Housing Prices, and Life-Cycle Events. Journal of Urban Planning and Development 141, 04014041. doi:\href{https://doi.org/10.1061/(ASCE)UP.1943-5444.0000257}{10.1061/(ASCE)UP.1943-5444.0000257}

\leavevmode\hypertarget{ref-RashidiModelinginterdependenciesvehicle2011}{}%
Rashidi, T.H., Mohammadian, A., Koppelman, F.S., 2011. Modeling interdependencies between vehicle transaction, residential relocation and job change. Transportation 38, 909.

\leavevmode\hypertarget{ref-rcoreteamLanguageEnvironmentStatistical2019}{}%
R Core Team, 2019. R: A language and environment for statistical computing. R Foundation for Statistical Computing, Vienna, Austria.

\leavevmode\hypertarget{ref-rephannEconomicDemographicEffectsImmigration2004}{}%
Rephann, T.J., Holm, E., 2004. Economic-Demographic Effects of Immigration: Results from a Dynamic Spatial Microsimulation Model. International Regional Science Review 27, 379--410. doi:\href{https://doi.org/10.1177/0160017604267628}{10.1177/0160017604267628}

\leavevmode\hypertarget{ref-rutterDynamicMicrosimulationModels2011}{}%
Rutter, C.M., Zaslavsky, A.M., Feuer, E.J., 2011. Dynamic Microsimulation Models for Health Outcomes: A Review. Medical Decision Making 31, 10--18. doi:\href{https://doi.org/10.1177/0272989X10369005}{10.1177/0272989X10369005}

\leavevmode\hypertarget{ref-SalviniILUTEOperationalPrototype2005}{}%
Salvini, P., Miller, E.J., 2005. ILUTE: An Operational Prototype of a Comprehensive Microsimulation Model of Urban Systems. Networks and Spatial Economics 5, 217--234. doi:\href{https://doi.org/10.1007/s11067-005-2630-5}{10.1007/s11067-005-2630-5}

\leavevmode\hypertarget{ref-summerfield2011hilda}{}%
Summerfield, M., Hahn, M., n.d. HILDA user manual--release 10.

\leavevmode\hypertarget{ref-templSimulationSyntheticComplex2017}{}%
Templ, M., Meindl, B., Kowarik, A., Dupriez, O., 2017. Simulation of Synthetic Complex Data: The R Package simPop. Journal of Statistical Software 79. doi:\href{https://doi.org/10.18637/jss.v079.i10}{10.18637/jss.v079.i10}

\leavevmode\hypertarget{ref-vansonsbeekDisabilityBenefitMicrosimulation2012}{}%
van Sonsbeek, J.-M., Alblas, R., 2012. Disability benefit microsimulation models in the Netherlands. Economic Modelling 29, 700--715. doi:\href{https://doi.org/10.1016/j.econmod.2012.01.004}{10.1016/j.econmod.2012.01.004}

\leavevmode\hypertarget{ref-wuMosesDynamicSpatial2012}{}%
Wu, B., Birkin, M., 2012. Moses: A Dynamic Spatial Microsimulation Model for Demographic Planning, in: Tanton, R., Edwards, K. (Eds.), Spatial Microsimulation: A Reference Guide for Users. Springer Netherlands, Dordrecht, pp. 171--193. doi:\href{https://doi.org/10.1007/978-94-007-4623-7_11}{10.1007/978-94-007-4623-7\_11}

\leavevmode\hypertarget{ref-yeMethodologyMatchDistributions2009}{}%
Ye, X., Konduri, K., Pendyala, R.M., Sana, B., Waddell, P., 2009. A methodology to match distributions of both household and person attributes in the generation of synthetic populations, in: 88th Annual Meeting of the Transportation Research Board, Washington, DC.

\leavevmode\hypertarget{ref-zaidiDynamicMicrosimulationModels2001}{}%
Zaidi, A., Rake, K., 2001. Dynamic microsimulation models: A review and some lessons for SAGE. Sage discussion paper.

\leavevmode\hypertarget{ref-zucchelli2010evaluation}{}%
Zucchelli, E., Jones, A.M., Rice, N., others, 2010. The evaluation of health policies through microsimulation methods. Health, Econometrics and Data Group (HEDG) Working Papers 10.
\end{cslreferences}

\end{document}